\documentclass{svjour3} 
\smartqed 
\usepackage[utf8]{inputenc}
\usepackage{amsmath}
\usepackage{amsfonts}
\usepackage{amssymb}
\usepackage{listings}
\usepackage{mathtools}
\usepackage{mathrsfs} 
\usepackage{times}
\usepackage{float}
\usepackage{color}
\usepackage{setspace}
\usepackage{color,soul}

\usepackage{amsmath,amsfonts,amsthm,eucal}
\usepackage{enumerate}
\usepackage[letterpaper,margin=3cm]{geometry}
\usepackage{float}
\usepackage[title]{appendix}
\usepackage{color}
\usepackage{graphics}
\usepackage{stringenc}
\usepackage[
pdfstartview=XYZ,
bookmarks=true,
colorlinks=true,
linkcolor=blue,
urlcolor=blue,
citecolor=blue,
pdftex,
bookmarks=true,
linktocpage=true,   
hyperindex=true
]{hyperref}
\usepackage{lipsum}
\usepackage[FIGBOTCAP,TABTOPCAP,bf,tight]{subfigure}
\usepackage{natbib}
\usepackage[pdftex]{graphicx}
\usepackage{epstopdf}
\usepackage{booktabs} 
\usepackage{lineno}
\usepackage{empheq}

\usepackage{bm,upgreek}
\usepackage{tikz,mathpazo}
\usetikzlibrary{shapes.geometric, arrows}

\usepackage{caption}

\journalname{Computer Methods in Applied Mechanics and Engineering}

\newcommand{\tensor}[1]{\ensuremath{\boldsymbol{#1}}}

\DeclareMathOperator{\grad}{\nabla}

\theoremstyle{remark}

\renewcommand{\vec}[1]{\ensuremath{\boldsymbol{#1}}}

\DeclareMathOperator*{\argmin}{arg\,min}

\usepackage{algorithm}
\usepackage[noend]{algpseudocode}
\usepackage{multirow}

\usepackage{CJKutf8}
\usepackage{kotex}

\usepackage{regexpatch}
\usepackage{listings}

\usepackage{xcolor}

\DeclareFixedFont{\ttb}{T1}{txtt}{bx}{n}{9} 
\DeclareFixedFont{\ttm}{T1}{txtt}{m}{n}{9}  

\usepackage{color}
\definecolor{deepblue}{rgb}{0,0,0.5}
\definecolor{deepred}{rgb}{0.6,0,0}
\definecolor{deepgreen}{rgb}{0,0.5,0}
\definecolor{terminalblack}{rgb}{0.25,0.25,0.25}
\definecolor{f77green}{rgb}{0,0.7,0}
\definecolor{f77blue}{rgb}{0.0,0,0.7}

\newcounter{python}

\newcounter{fortran}

\makeatletter

\lst@UserCommand\lstlistofpython{\bgroup
    \let\contentsname\lstlistpythonname
    \let\lst@temp\@starttoc \def\@starttoc##1{\lst@temp{lop}}%
    \tableofcontents \egroup}
\lstnewenvironment{python}[1][]{%
  \let\c@lstlisting=\c@python
  
  \xpatchcmd*{\lst@MakeCaption}{lol}{lop}{}{}%
  \lstset{language=Python,
  basicstyle=\small,
  otherkeywords={self},
  tabsize=1,
  keywordstyle=\ttb\color{deepblue},
  emph={__init__, update_port},
  emphstyle=\ttb\color{deepred},
  stringstyle=\ttb\color{deepgreen},
  commentstyle=\color{f77green},
  frame=single,
  showstringspaces=false,
  float=htpb,
  captionpos=b,
  numbersep=5pt,
  linewidth=0.99\linewidth,
  xleftmargin=0.01\linewidth,
  breaklines=true
  breakwhitespace=false,
  #1}}
  {}

\lst@UserCommand\lstlistoffortran{\bgroup
    \let\contentsname\lstlistfortranname
    \let\lst@temp\@starttoc \def\@starttoc##1{\lst@temp{lof}}%
    \tableofcontents \egroup}
\lstnewenvironment{fortran}[1][]{%
  \let\c@lstlisting=\c@fortran
  
  \xpatchcmd*{\lst@MakeCaption}{lol}{lof}{}{}%
  \lstset{language=[90]fortran,
  basicstyle=\ttfamily\small,
  showstringspaces=false,
  keywordstyle=\color{f77blue},
  stringstyle=\color{purple},
  commentstyle=\color{f77green},
  breakatwhitespace=false,
  frame=single,
  linewidth=0.99\linewidth,
  xleftmargin=0.01\linewidth,
  breaklines=true
  breakwhitespace=false,
  showstringspaces=false,
  captionpos=b,
  #1}}
  {}

\usepackage{algorithm}
\usepackage{algpseudocode}

\algdef{SE}[SUBALG]{Indent}{EndIndent}{}{\algorithmicend\ }%
\algtext*{Indent}
\algtext*{EndIndent}

\title{
Discovering interpretable elastoplasticity models via the neural polynomial method enabled symbolic regressions
}

\begin{document}

\titlerunning{Interpretable ML plasticity}

\author{Bahador Bahmani \and Hyoung Suk Suh \and WaiChing Sun}

\institute{Corresponding author: WaiChing Sun 
\at Associate Professor,
Department of Civil Engineering and Engineering Mechanics, Columbia University,
614 SW Mudd, Mail Code: 4709, New York, NY 10027 Tel.: 212-854-3143, Fax:
212-854-6267, 
  \email{wsun@columbia.edu}
}

\date{Received: \today / Accepted: date}

\maketitle

\begin{abstract}
Conventional neural network elastoplasticity models are often perceived as lacking interpretability. 
 This paper introduces a two-step machine learning approach that returns mathematical models interpretable by human experts.  In particular, we introduce a surrogate model where yield surfaces are expressed in terms of a set of single-variable feature mappings obtained from supervised learning.  A post-processing step is then used to re-interpret the set of single-variable neural network mapping functions into mathematical form through symbolic regression.  This divide-and-conquer approach provides several important advantages.  First,  it enables us to overcome the scaling issue of  symbolic regression algorithms. From a practical perspective,  it enhances the portability of learned models for partial differential equation solvers written in different programming languages.  Finally, it enables us to have a concrete understanding of the attributes of the materials, such as convexity and symmetries of models,  through automated derivations and reasoning. Numerical examples have been provided, along with an open-source code to enable third-party validation.

\end{abstract}


\section{Introduction}
\label{sec:intro}
In the last decade, the number of machine learning constitutive models has increased significantly \citet{ghaboussi1991knowledge,  pernot1999application,   mozaffar2019deep, logarzo2021smart,  liu2021review}. Among those machine learning models, neural networks trained with experimental or simulation data have been 
one of the most popular choices \citep{wang2018multiscale,  vlassis2020geometric,  vlassis2021sobolev, flaschel2022discovering}.
Despite the recent popularity of these neural network models and a few attempts to adapt machine learning models into production software
 \citet{li2019machine,  suh2023publicly},
the adaptation of these constitutive models to high-consequence engineering applications has not yet been mainstream.
Potential issues could be attributed to the lack of reproducibility of the neural network models (cf. \citet{suh2023publicly}),  insufficient interpretability/explainability \citep{fan2021interpretability,  murdoch2019definitions},  the difficulty of striking the balance between accuracy and robustness  (cf.  \citet{raghunathan2020understanding,  sagawa2019distributionally}),
and a combination of these issues that makes the trustworthiness of the trained model questionable  \citet{wing2021trustworthy}.

There have been attempts to improve the interpretability of the machine learning models with different degrees of success. 
 \citet{vlassis2021sobolev} and \citet{vlassis2022component},  for instance,  introduce a component-based design for neural network plasticity models.  
This approach trains separated neural network models for hyperelastic stored energy functionals and yield surfaces with hardening laws. 
As such,  geometrical features of those learned functions,  such as convexity (and the lack thereof) of the elastic energy functional and the non-smoothness and symmetry of plastic flow, 
can be interpreted and correspond with the specific properties of the materials,  such as material stability,   phase transition,  and the existence of discrete mechanisms (e.g. , a slip system) and material symmetry.  The similar component-based idea has been incorporated in a modular design machine learning framework for elastoplasticity \citep{fuhg2023modular}. 
In both cases,  this geometrical interpretation is a departure from the recurrent neural network approach or multi-step feedforward neural network approaches 
where the yield surface is not explicitly defined, but could be recovered in a post-hoc analysis,  as shown in \citep{mozaffar2019deep}. 
Other related efforts to introduce more interpretable models include the incorporation of knowledge graphs 
\citep{ wang2019meta,  he2022thermodynamically} and causal discovery for constitutive responses \citep{sun2022data}. 
These graph-based approaches may provide relational and structural knowledge about the learned material models. 
The relations represented by graphs can then be interpreted as falsifiable propositions (and/or hypotheses) and thus enable easier third-party scrutiny and inspections. 
Furthermore,  post-hoc analysis can be an alternative approach to interpret models.  
For instance,  one may test (through random sampling or adversarial attacks (cf. \citep{wang2021non})  whether the learned model holds 
the necessary properties of the ground truth.
These necessary properties can be universal principles,  such as thermodynamic laws,  
or prior knowledge of material behaviors, such as material symmetry due to crystal structures and convexity of energy functionals due to the observed stable behaviors \citep{vlassis2021sobolev}.  However,  tests based on samplings alone are insufficient to provide definite proof of propositions. 

Another approach to enhance the interpretability of elastoplasticity models is to perform symbolic regressions directly to learn a portion or all of the plasticity models \citep{versino2017data,  wang2022establish,  bomarito2021development}.  
The advantage of this approach is that it may lead to a mathematical expression of the learned function that is much shorter than the neural network counterparts and,  hence,  suitable for analysis and reduces the execution time of the constitutive laws \cite{suh2023publicly}. 
However,  as symbolic regression requires solving combinatorial optimizations to find the optimal equation expressed as an expression tree,  the number of possible combinations of symbolic expression grows rapidly with the dimensionality of the input and output.  It is an NP-hard problem (cf. \citet{mundhenk2021symbolic}) where even the state-of-the-art symbolic regression algorithm exhibits known difficulty in interpolating multi-dimensional functions \citep{petersen2019deep}. 

On a related note,  \citet{linka2023new,  linka2023automated},  and \citet{tacc2023benchmarking} apply a different symbolic regression approach in which a set of prior hyperelasticity models are chosen as the basis functions for biological tissues.  An optimization problem is then solved to determine the coefficients of the learned models. In principle,  this interpolation technique can also be used for learning yield functions or hardening laws.   Since such a model is a linear combination of the 
established hand-crafted models,  the resultant models can be perceived as easier to interpret. 

However,  the accuracy of the resultant model could be jeopardized if the basis models do not span a finite dimensional space that yields good fitting.  For instance,  it is not possible to capture a pressure-sensitive yielding (e.g. ,  Drucker-Prager model) well by using yield functions expressed on the $\pi-$plane (e.g.,  von Mises and Tersca models) as the basis functions (although the least square solution that averages the pressure-sensitivity effect along the hydrostatic axis can be found).  
Presumably,  this issue can be alleviated by increasing the number of basis models to improve the expressivity \footnote{The expressivity of a neural network architecture refers to the size of the set (or cardinality) of all possible functions a specific neural network architecture is capable of approximating (cf.  \citep{raghu2017expressive,lin2018generalization}).  The expressivity is a necessary but not sufficient condition for accurate model }.  However,  this may require good prior knowledge and intuitions of the relationship between the data set and the basis models.  The numerical stability of the optimization problem may also require independence of the basis models to ensure the uniqueness of the coefficients/weights.

\begin{figure}[h!]
\centering
\includegraphics[height=0.45\textwidth]{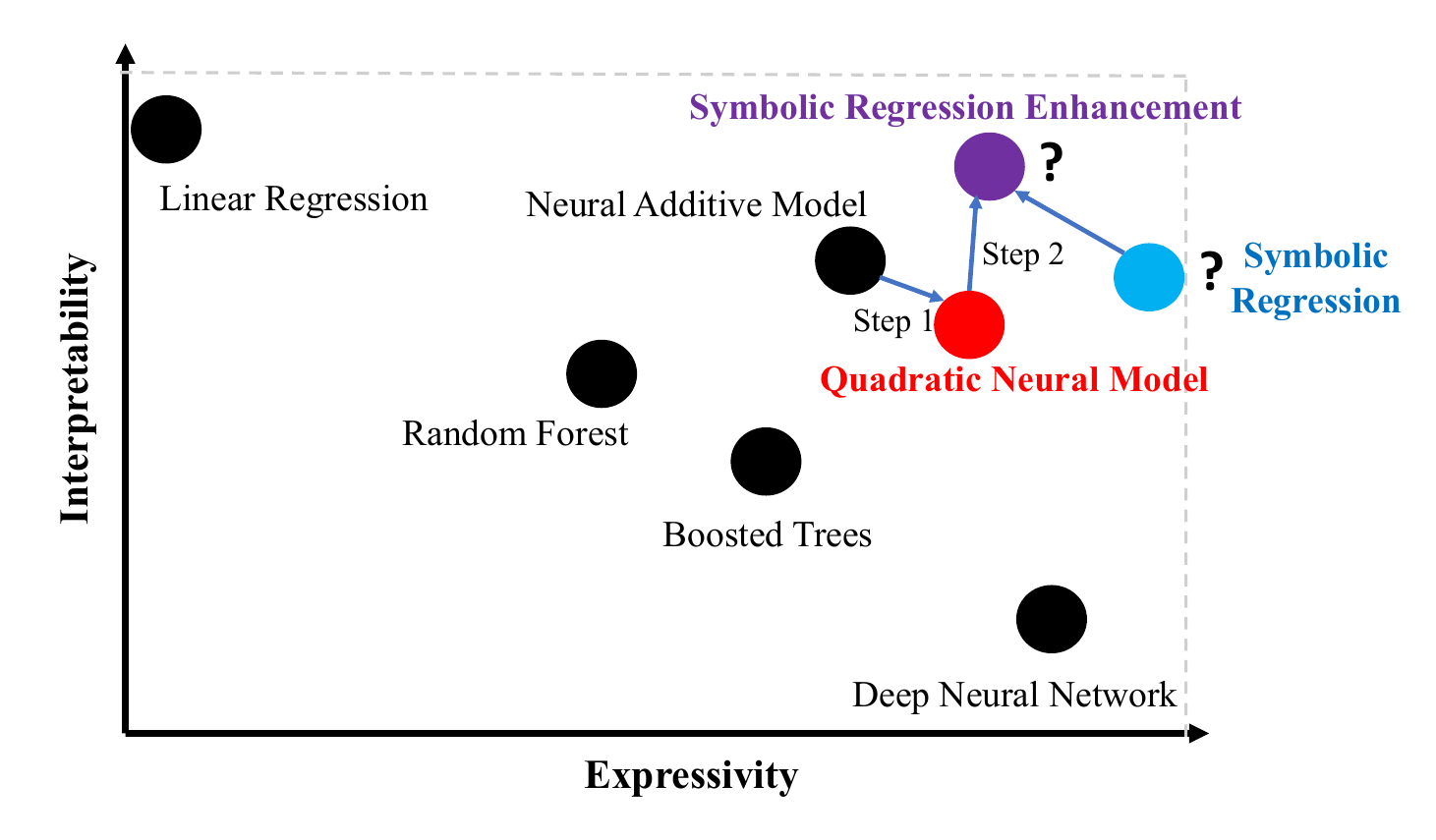}
\caption{
 The trade-off between expressivity and interpretability in various machine learning models. 
We introduced the Quadratic Neural Model (QNM), which enhances the expressivity of the Neural Additive Model (NAM) 
at the expense of reducing interpretability.
However, by obtaining an analytical expression of the feature space mapping,  we can achieve a better trade-off between expressivity and interpretability.}
\label{fig:tradeoff}
\end{figure}

As demonstrated in the literature,  achieving the optimized trade-off among expressivity,  generalization,  trainability,  execution speed, and interpretability remains a challenging problem for machine learning constitutive laws. 

\subsection{Interpretability vs. expressivity}
Multilayer perceptrons (MLP) have been demonstrated to be good candidates for supervised regression tasks where a single fully-connected deep neural network
is utilized as the model class \citep{szegedy2013intriguing}.  
MLP is also robust in learning nonlinear models that can distinguish data that is not linearly separable.
Due to the high expressivity of deep neural networks and the inherent nonlinear interactions among the input features,  a consistently high accuracy has been reported by those successfully trained MLP models. 
However, it could become challenging to interpret/extract the input-output relationships for neural network models with multivariate inputs because they may combine the input features in a highly nonlinear manner; this entanglement makes it difficult to isolate the effect of each feature on the output \citep{peng2019domain}. 

This black-box issue is not only a technical barrier to computational
 mechanics but also critical for many other disciplines, such as 
 drug delivery,  where the interpretability of the solution is critical.
\citet{doran2017does}, for instance,  define interpretable systems: \textit{“A system where a user cannot only see, but also study and understand how inputs are mathematically mapped to outputs."}
\citet{gilpin2018explaining} define intelligibility as a combination of explainability (being able to provide a rationale for the results, sometimes through posthoc analysis) and interpretability (the logic that delivers the learned results can be comprehended by humans.)
In many cases,  explainability could be achieved by 
model-agnostic methods developed to explain the predictions of black-box models via the feature importance and local approximation, as pointed out by \citet{xu2022sparse}.  
Meanwhile, models that are inherently interpretable, such as decision-tree-based models, often introduce mechanisms (e.g.,  hierarchical decisions or rules) such that the rationale of the trained model can be understood. 

Nevertheless, as pointed out by \citet{agarwal2021neural},  
machine learning techniques that exhibit high interpretability 
often lack the level of expressivity (the ability to express an arbitrary function -- a necessary but not sufficient condition for accuracy) to yield accurate predictions for complex tasks. 
Fig. \ref{fig:tradeoff} (modified from \citet{agarwal2021neural}) illustrates the trade-off between interpretability and expressivity for a variety of common machine learning models,  which include deep neural network,  boosted tree,  and random forest. 
As supported by the universal approximation theorem \citep{hornik1989multilayer}, the deep neural network is often considered a machine learning tool with high expressivity but also 
difficult to interpret. Meanwhile,  linear regression is easy to interpret but often lacks the expressivity for more complex tasks.

In theory,  a symbolic regression,  if conducted successfully,  may achieve both the desirable level of expressivity and interpretability if the combinatorial optimization that searches the optimal expression tree is successful. 
However, the symbolic regression problem, in particular for high dimensional data,  is an NP-hard problem (cf. \citet{udrescu2020ai}). Hence, it is difficult to ensure the training performance of 
symbolic regression or to estimate the probability of success for the multi-dimensional data. 
An interesting benchmark study has been conducted by 
\citet{petersen2019deep} using 6 state-of-the-art symbolic regression software packages.
While these software packages demonstrate a degree of success in univariate function, they all failed 
 in recovering a fourth-order polynomial with two variables (see Table \ref{Tab:peterson}), i.e., 
\begin{equation}
f(x,y) = x^4 - x^3 + \frac{1}{2}y^{2} - y.
\label{eq:difficultequation} 
\end{equation}

Hence,  a strategy that effectively addresses the NP-hard symbolic regression problem for high-dimensional contexts without comprising expressivity could be a breakthrough not only for mechanics but could also be significant for advancing state-of-the-art symbolic regression through approximations,  heuristics,  or specialized algorithms. 

\begin{table}
\begin{tabular}{cccccccc} 
Benchmark & Expression & DSR & PQT & VPG & GP & Eureqa & Wolfram \\
\hline Nguyen-1 & $x^3+x^2+x$ & $100 \%$ & $100 \%$ & $96 \%$ & $100 \%$ & $100 \%$ & $100 \%$ \\
Nguyen-2 & $x^4+x^3+x^2+x$ & $100 \%$ & $99 \%$ & $47 \%$ & $97 \%$ & $100 \%$ & $100 \%$ \\
Nguyen-3 & $x^5+x^4+x^3+x^2+x$ & $100 \%$ & $86 \%$ & $4 \%$ & $100 \%$ & $95 \%$ & $100 \%$ \\
Nguyen-4 & $x^6+x^5+x^4+x^3+x^2+x$ & $100 \%$ & $93 \%$ & $1 \%$ & $100 \%$ & $70 \%$ & $100 \%$ \\
Nguyen-5 & $\sin \left(x^2\right) \cos (x)-1$ & $72 \%$ & $73 \%$ & $5 \%$ & $45 \%$ & $73 \%$ & $2 \%$ \\
Nguyen-6 & $\sin (x)+\sin \left(x+x^2\right)$ & $100 \%$ & $98 \%$ & $100 \%$ & $91 \%$ & $100 \%$ & $1 \%$ \\
Nguyen-7 & $\log (x+1)+\log \left(x^2+1\right)$ & $35 \%$ & $41 \%$ & $3 \%$ & $0 \%$ & $85 \%$ & $0 \%$ \\
Nguyen-8 & $\sqrt{x}$ & $96 \%$ & $21 \%$ & $5 \%$ & $5 \%$ & $0 \%$ & $71 \%$ \\
Nguyen-9 & $\sin (x)+\sin \left(y^2\right)$ & $100 \%$ & $100 \%$ & $100 \%$ & $100 \%$ & $100 \%$ & - \\
Nguyen-10 & $2 \sin (x) \cos (y)$ & $100 \%$ & $91 \%$ & $99 \%$ & $76 \%$ & $64 \%$ & - \\
Nguyen-11 & $x^y$ & $100 \%$ & $100 \%$ & $100 \%$ & $7 \%$ & $100 \%$ & - \\
Nguyen-12 & $x^4-x^3+\frac{1}{2} y^2-y$ & $0 \%$ & $0 \%$ & $0 \%$ & $0 \%$ & $0 \%$ & - \\
\cline { 3 - 7 } & Average & $\mathbf{8 3 . 6 \%}$ & $75.2 \%$ & $46.7 \%$ & $60.1 \%$ & $73.9 \%$ & -
\end{tabular}
\caption{
\label{Tab:peterson}
Comparison of symbolic equation recovery rate among various symbolic regression algorithms implemented in different packages.  This table is reproduced from \cite{petersen2019deep}; please refer to Table 1 in the mentioned reference for more details.
}
\end{table}

\subsection{Neural Additive Models: trade-off for interpretability and expressivity with linear feature space}
\label{sec:nam}
\citet{agarwal2021neural} propose the Neural Additive Model (NAM) in which a  set of independent neural networks are co-trained to generate a set of nonlinear scalar features $f_{i}(x_{i})$, one for each input $x_{i}$ where $i=1,2,...,D$ and $D$ is the number of input dimensions.  The model structure is the linear combination of these scalar features, i.e., 
\begin{equation}
\bar{\phi}(\vec{x}; \vec{\beta},  \vec{\omega}) = \sum_{i=1}^D w_{i} f_i(x_i; \vec{\beta}_i),
\label{eq::nam}
\end{equation}
where each feature function $f_i$ is parameterized by a multilayer perceptron (MLP) with parameters $\vec{\beta}_i \in \mathbb{R}^{M_i}$,  $M_i$ is the total number of trainable parameters of the \textit{i}-th MLP,  and $\vec{w} \in \mathbb{R}^D$.
These single-variable MLP functions are referred to as \textbf{shape (basis) functions}.
The contribution of each shape function is controlled by the trainable (weighting) parameters $w_{i} \in \mathbb{R}$.  The vector $\vec{\beta} = \{ \vec{\beta}_i \}_{i=1}^D$ concatenates all neural network related parameters (weights and biases).

\citet{agarwal2021neural} argue that this approach is interpretable in the sense that the importance of each feature 
can be ranked by examining the coefficients of the feature $w_{i}$. In other words,  
the NAM approach maintains the interpretability of the linear regression (in the feature space) with enhanced expressivity afforded by the neural networks. 
However, \citet{agarwal2021neural} also point out that NAM
exhibits less expressivity of the fully connected neural network, 
especially when expressing the ground-truth function requires
bases independent of the feature basis functions.

\subsection{Proposed strategy for interpretable model recovery}
Given the fact that constitutive laws are often used for high-consequence engineering applications,  
making the machine learning generated constitutive laws interpretable is necessary (but not sufficient) to ensure trustworthiness.
The purpose of this research is to propose a new supervised machine learning method for elastoplasticity models that strikes a balance between expressivity and interpretability (see Fig. \ref{fig:tradeoff}).  
To achieve this objective,  we first take the neural network architecture of the neural additive model proposed by \citet{agarwal2021neural} (which was originally designed to achieve interpretability through linear regression of feature space generated by neural networks),  and make two major modifications of the supervised learning problem.  First,  
we improve the expressivity of the learned models by generalizing the feature space to be a polynomial of univariate functions learned from data. 
Expressing the model as a polynomial in the feature space enables us to improve expressivity systematically at a known expense of increased complexity of the resultant models (see Section \ref{sec:qnm}). 
Second,  instead of relying on the coefficient of the feature basis to determine importance,  we intend to improve interpretability 
by re-expressing the set of univariate functions in symbolic form.   In particular, we leverage the fact that the feature space is spanned by univariate functions. 

This setting enables us to break down the  NP-hard high-dimensional symbolic regression problem into a series of separated one-dimensional symbolic regressions (see Section \ref{sec:symb-regr}), which have consistently been successful in discovering yield surfaces and the underlying hardening mechanisms in our numerical experiments (see Section \ref{sec:num_example}.)

One should be cautious against the expectation that the proposed method should \textbf{robustly} recover the exact mathematical expression used to generate the data for the following reasons. First, many mathematical equations can be \textbf{expressed} in multiple equivalent ways (e.g. ,  
$\sin(x) = \cos(\pi/2 - x) = (e^{ix} - e^{-ix})/2i$ ).  
Although these equations possess the same information,  they are represented by different expression trees,  as shown in Section \ref{sec:symb-regr}.   This non-uniqueness makes finding the identical expression of the benchmark solution (if available at all) more difficult, given the curse of high-dimensionality of the underlying combinatoric optimization problem one must solve.

Secondly,  from a practical perspective,  it is debatable whether assuming the existence of a ground-truth mathematical form of the 
yield surfaces,  which is often derived to match macroscopic phenomenological behaviors with physics justifications (e.g., von Mises,  Drucker-Prager,  Mohr-Coulomb yield models) is necessary.   \citet{dafalias2004simple} and \citet{dafalias202113},  for example,  argue that the choices of the mathematical formula to model the same materials is often a trade-off between accuracy and simplicity.  Similar assertions can also be found in a large-scale symbolic regression benchmark study conducted by \citep{la2021contemporary},  where the authors argue that a model should be regarded as a symbolic solution to a problem with ground-truth solution that generates the data  if the learned model depends on the same variables and either the discrepancy or ratio between the model and the ground-truth is finite (see Definition 4.1 in  \citep{la2021contemporary}).

\section{Method}
\label{sec:interp_ML}

We begin by introducing the general problem statement of finding yield surface as a supervised regression problem in Section \ref{sec:prob_state}. We then introduce the quadratic neural model (QNM) in Section \ref{sec:qnm}.  We explain our choice of neural network architecture in Section \ref{sec:nn-model}. Additionally, in Section \ref{sec:sparse-loss}, we specify a sparsity-promoting constraint used during the training process to enhance the simplicity and interpretability of the model.  For completeness, we provide details of the genetic programming algorithm that conducts the symbolic regression for the feature bases in Section \ref{sec:symb-regr}.

\subsection{Problem statement}
\label{sec:prob_state}
Our learning task is to find a mapping function $\phi(\vec{x}): \mathbb{R}^D \to \mathbb{R}$ from  any element of $D$-dimensional Euclidean space onto real numbers, where $\vec{x}$ 
is the state variable for the yield function, including the Cauchy stress and internal variables, $D$ is the dimension of the inputs,  and $\phi(\vec{x})$ is the yield function. 
Given $N$ data points stored as a point cloud $\mathcal{C} = \{ \vec{x}^l,  \phi^l \}_{l=1}^{N}$,  we approximate such a function by the parametric function $\bar{\phi}(\vec{x};\vec{\beta}, \vec{w})$, e.g.,  Equation \eqref{eq::nam}, where the best estimator for its unknown parameters $\vec{\beta}$ and $\vec{w}$ are found via,  in the sense of least square,
\begin{equation}
	\vec{\beta},  \vec{w} = \underset{\vec{\beta}, \vec{w}}{\argmin} 
	\frac{1}{N}
	\sum_{l=1}^N 
	(\bar{\phi}(\vec{x}^l; \vec{\beta}, \vec{w}) - \phi^l)^2
	+
	\mathcal{L}_{\text{sparsity}}(\vec{w}),
\label{eq::prob-state}
\end{equation}
where $\mathcal{L}_{\text{sparsity}}$ is a regularization term for sparsity control (see Section \ref{sec:sparse-loss}).
In the conventional setting,  one may use a multivariate fully connected neural network as the model class.  A major departure is that we will instead postulate the existence of a feature space spanned by basis functions $f_{i}$ learned by univariate neural networks (see Section \ref{sec:qnm}). 
This configuration enables us to obtain sufficiently expressive models $\bar{\phi}(\vec{x}^{l})$  while interpretability is guaranteed via symbolic regressions that replace the trained neural networks that parametrize basis functions of the feature space.

\subsection{Quadratic Neural Model for enhanced expressivity}
\label{sec:qnm}
\begin{figure}[h!]
\centering
{\includegraphics[width=0.95\textwidth]{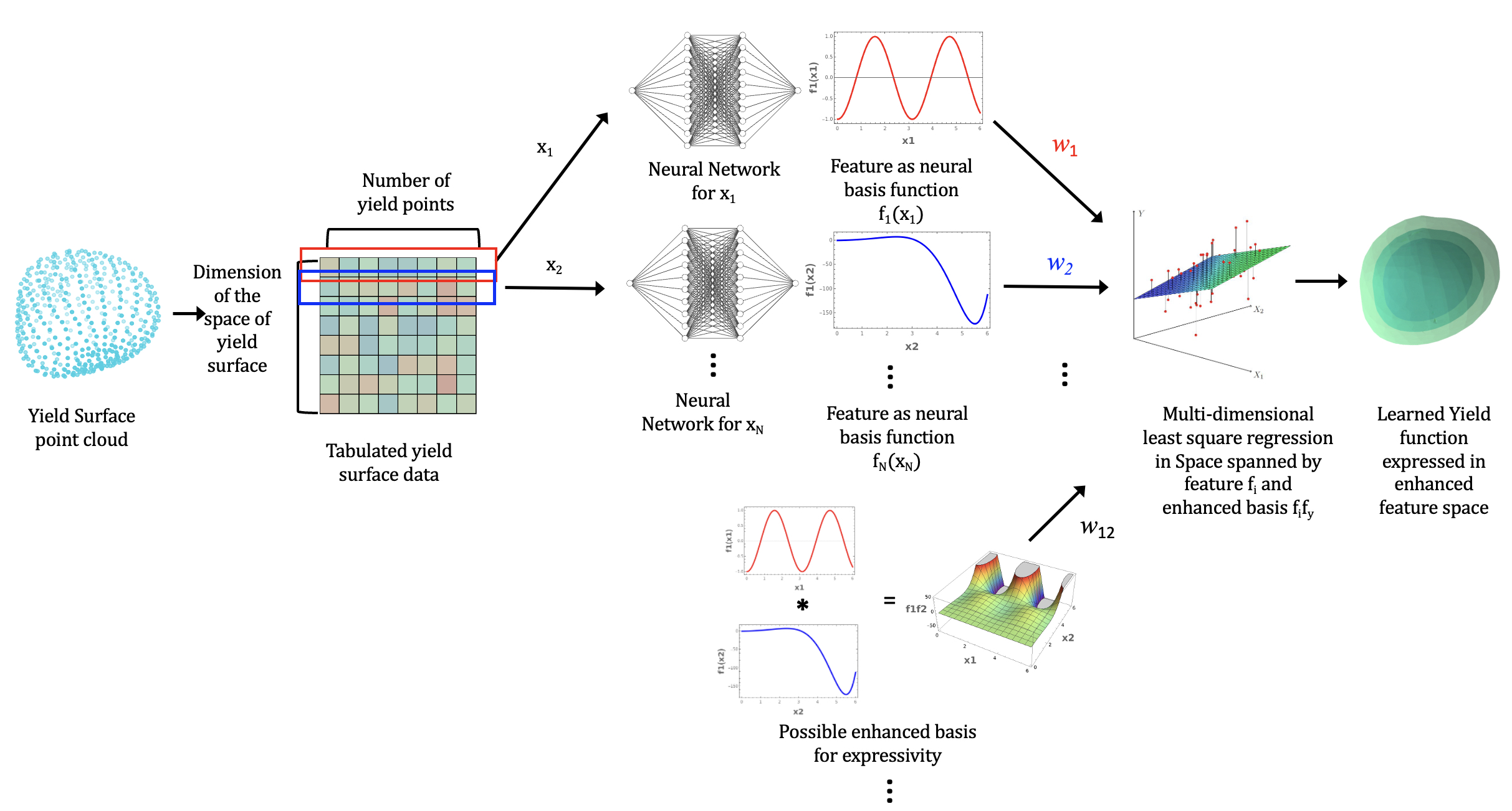}}
\caption{The neural quadratic method for enhanced expressibility. Instead of using a fully connected neural network with a multi-dimensional input layer, the proposed univariate neural networks are trained to create feature space, forming the basis to express the yield function analytically. To enhance expressivity, additional bases founded by the product of features are introduced.}
\label{fig:step1}
\end{figure}

As mentioned in Section \ref{sec:nam}, the original neural additive model enhances the interpretability of the learned model through the generation of feature basis. The resultant model then becomes linear in the feature space. This enhanced interpretability is, nevertheless, achieved at the expense of expressivity. To circumvent
this limitation, we generalize the formulation of the neural additive model to introduce additional quadratic terms (see Fig. ~\ref{fig:step1}.) As such, we refer to this revised approach as the Quadratic Neural Model (QNM.)

The resultant learned yield function expressed with the additional enhancement bases reads:
\begin{equation}
\bar{\phi}(\vec{x}; \vec{\beta},  \vec{\omega}) = \sum_{i=1}^D w_{i} f_i(x_i; \vec{\beta}_i)
+
\sum_{i=1}^D \sum_{j=i}^D \hat{w}_{ij} f_i(x_i; \vec{\beta}_i) f_j(x_j; \vec{\beta}_j), 
\label{eq::qnm}
\end{equation}
where $\hat{w}_{ij}$ are additional trainable parameters that control contributions of the second-order interactions between $\textit{i}$-th and $\textit{j}$-th shape functions.  
$\hat{\vec{w}} \in \mathbb{R}^{D + D (D-1)/2}$ denotes the concatenation of all parameters $\vec{w} = \{ w_i,  \hat{w}_{ij}\} $.
Notice that the same functions as those used in the first-order term are utilized for the second-order term; in total, there are $D$ numbers of different shape functions to be learned simultaneously.

\subsubsection{Neural network architecture for shape (basis) functions}
\label{sec:nn-model}

With the supervised learning problem defined,  our next goal is to learn both (1) the weights and bias ($\vec{\beta}_{i}$) of each feature univariate neural network that parametrizes individual shape (basis) function $f_i(x_i)$ where $i=1, 2, ..., D$ in Eq.  \eqref{eq::qnm} ,  and (2) the coefficients of the quadratic feature space $\vec{w}$.   As in the case of the NAM models,  this set of basis functions that spans the quadratic feature space is not pre-determined a priori but manifested from the training data through solving Eq.  \eqref{eq::prob-state}.   To ensure the expressivity of the univariate neural network shape function, we must first avoid the well-known 'lazy learning' behavior, which is the inability or slow learning of capturing high-frequency content, commonly exhibited in neural networks with a low number of input dimensions (cf. \citet{tancik2020fourier,rahaman2019spectral}).

In our work, we build upon recent developments by enriching classical neural networks with Fourier layers,  as described in \citep{rahimi2007random,tancik2020fourier}.  This architecture can capture high-frequency content and improve the overall performance of our model.

\begin{figure}[h!]
\centering
\includegraphics[height=0.35\textwidth]{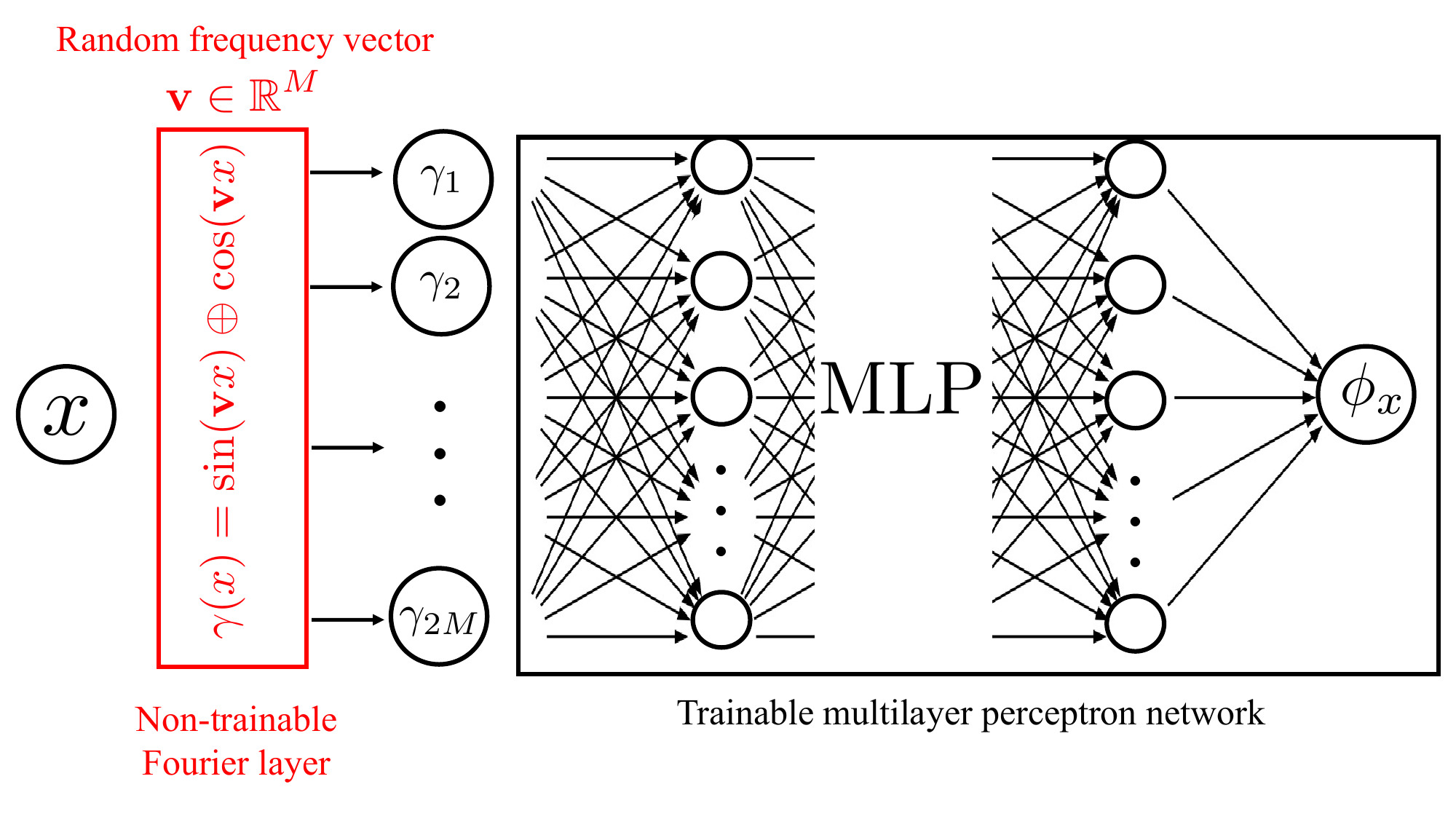}
\caption{Neural network architecture for each shape function.  A Fourier layer is utilized to improve the training of classical MLP.}
\label{fig:nn-arch}
\end{figure}

Each basis function $f_i(x_i)$ shown in Fig.~ \ref{fig:nn-arch} is parameterized by an MLP enriched with the Fourier layer as follows:
%
\begin{equation}
f_i(x_i; \vec{\beta}_i) =
h(
\vec{W}_L^i
\cdots
g(
\vec{W}_2^i
g(
\vec{W}_1^i 
\vec{\gamma}(x_i)
+ \vec{b}_1^i
)
+ \vec{b}_2^i
)
\cdots
+
\vec{b}_L^i
),
\end{equation}
where $\vec{W}_k^i$ and $\vec{b}_k^i$ are the weight and bias of the $k$-th hidden layer and $g(\cdot)$ and $h(\cdot)$ are hidden and output activation functions, respectively.  In this MLP function, the input layer $\vec{\gamma}(x_i)$ is the Fourier mapping of the input feature $x_i$ with random frequency vector $\vec{v} \in \mathbb{R}^{M}$ as follows:
\begin{equation}
\vec{\gamma}(x_i)
= 
\sin(\vec{v} x_i) 
\oplus
\cos(\vec{v} x_i)
=
[\sin(\vec{v} x_i)^T, \cos(\vec{v} x_i)^T]^T \in \mathbb{R}^{2M},
\end{equation}
where $\oplus$ indicates vector concatenation operation, and integer $M$ is an additional hyperparameter that indicates the number of hidden units in the Fourier layer.
Components of the random vector $\vec{v}$ are sampled from zero-mean normal distributional with standard deviation $\sigma_v$, i.e., $v_m \sim \mathcal{N}(0, \sigma_v)$. 
In this work, we keep the random Fourier features fixed during training,  although they can be considered trainable parameters.  
Previous research has demonstrated that optimizing them may not improve the approximation power and may increase the computational cost.  
All trainable parameters associated with the $i$-th shape function are denoted by
$\vec{\beta}_i = 
\{ 
\vec{W}_l, \vec{b}_l
\}_{l=1}^{L}$ where $L$ is the number of hidden layers.

\begin{figure}[h!]
\centering
\subfigure[]
{\includegraphics[height=0.375\textwidth]{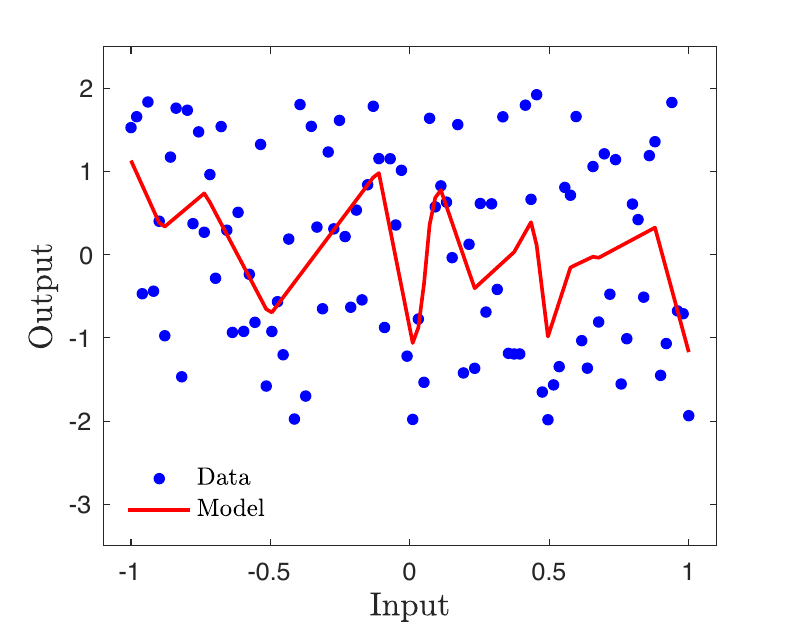}}
\subfigure[]
{\includegraphics[height=0.375\textwidth]{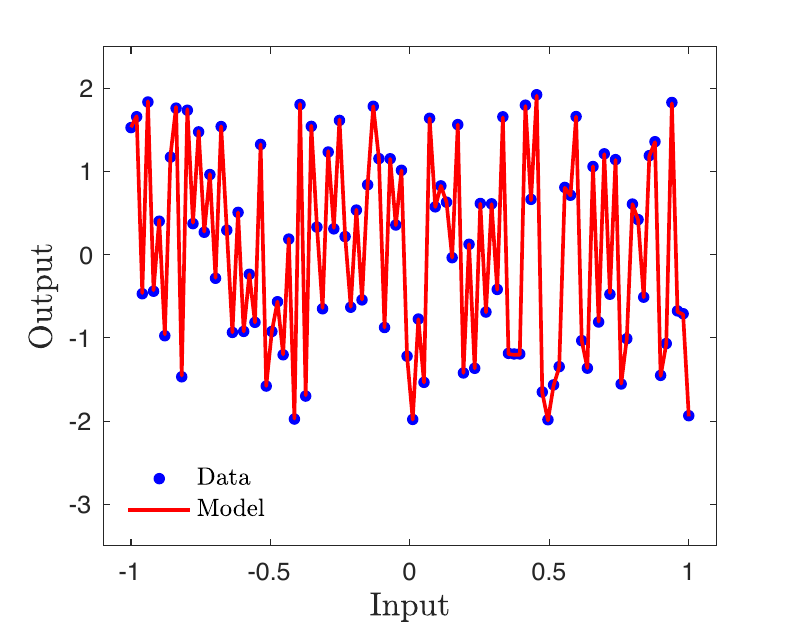}}
\caption{Network expressivity: (a) vanilla single-layer MLP with 80 hidden neurons and (b) single-layer spectral layer with 40 hidden neurons.  Both models are trained for 10,000 epochs with ADAM optimizer.  In this demonstration example, the 
data is intentionally over-fitted to test the expressivity power
of the spectral layer for complicated data.
}
\label{fig:noise-pred}
\end{figure}

To demonstrate the effectiveness of our architecture choice, we conducted an educational example following \citep{agarwal2021neural}. 
We generated a training dataset with high-frequency content using purely random noise and then empirically examined whether our architecture choice could overfit the training data. The results shown in Fig.~\ref{fig:noise-pred}, confirm that the neural network enriched with the spectral layer is generally capable of fitting highly complex data. We note that overfitting is generally not desirable, but in this particular example, our goal was to measure the expressivity and flexibility of the architecture by its ability to memorize the entire training dataset.

In this framework,  the contribution of each basis (shape) function $f_i$ in Eq.~ \ref{eq::qnm} is balanced by the associated weight $w_i$.  For example,  $w_i \gg w_j$ is intended to mean that the $i$-th shape function effect is much more important than the $j$-th shape function in the final prediction.  The necessary condition for this argument to be meaningful is that shape functions should have the same scale.  To achieve this, we apply the \texttt{tanh} activation function in the last layer of the neural network architecture,  which restricts the output to a range between -1 and 1.

\subsubsection{Regularization of polynomial function in feature space}
\label{sec:sparse-loss}
If ease of interpretation is the highest priority, one can set the yield function as a linear combination of shape functions. To further simplify the model and reduce dimensionality, one could also perform an additional feature extraction task, i.e., eliminating the shape function with the smallest coefficient, thereby removing features that do not significantly contribute to the accuracy of the approximation.

To incorporate regularization that favors interpretability and simplicity,  we include an additional term in the optimization statement Eq.~\ref{eq::prob-state} to promote sparsity for low-order and high-order terms. 
The $L0$-norm, which simply counts the total number of nonzero elements of a vector, is known as one of the best sparsity measures \citep{natarajan1995sparse,gale2019state}.  
 However, $L0$-norm minimization is an NP-hard problem and makes the proposed loss function in Eq.~\ref{eq::prob-state} non-differentiable \citep{natarajan1995sparse},  which is not preferred. 
As a result,  we use the $L1$-norm as a differentiable replacement of the $L0$-norm as follows \citep{tibshirani1996regression,brunton2016discovering}, 
\begin{equation}
\mathcal{L}_{\text{sparsity}}(\vec{w}) = 
\alpha_{\text{lo}} \sum_{i} |w_{i}| 
+ 
\alpha_{\text{ho}} \sum_{i,j} |w_{ij}|,
\label{eq::sparse_loss}
\end{equation}
where $\alpha_{lo}$ and $\alpha_{ho}$ are non-negative real-valued parameters that are predefined as hyperparameters to control the sparsity of both low-order and higher-order terms, with higher values imposing a stronger penalty.

\subsection{Symbolic Regression of feature space for enhanced interpretability}
\label{sec:symb-regr}

\begin{figure}[h!]
\centering
{\includegraphics[width=0.95\textwidth]{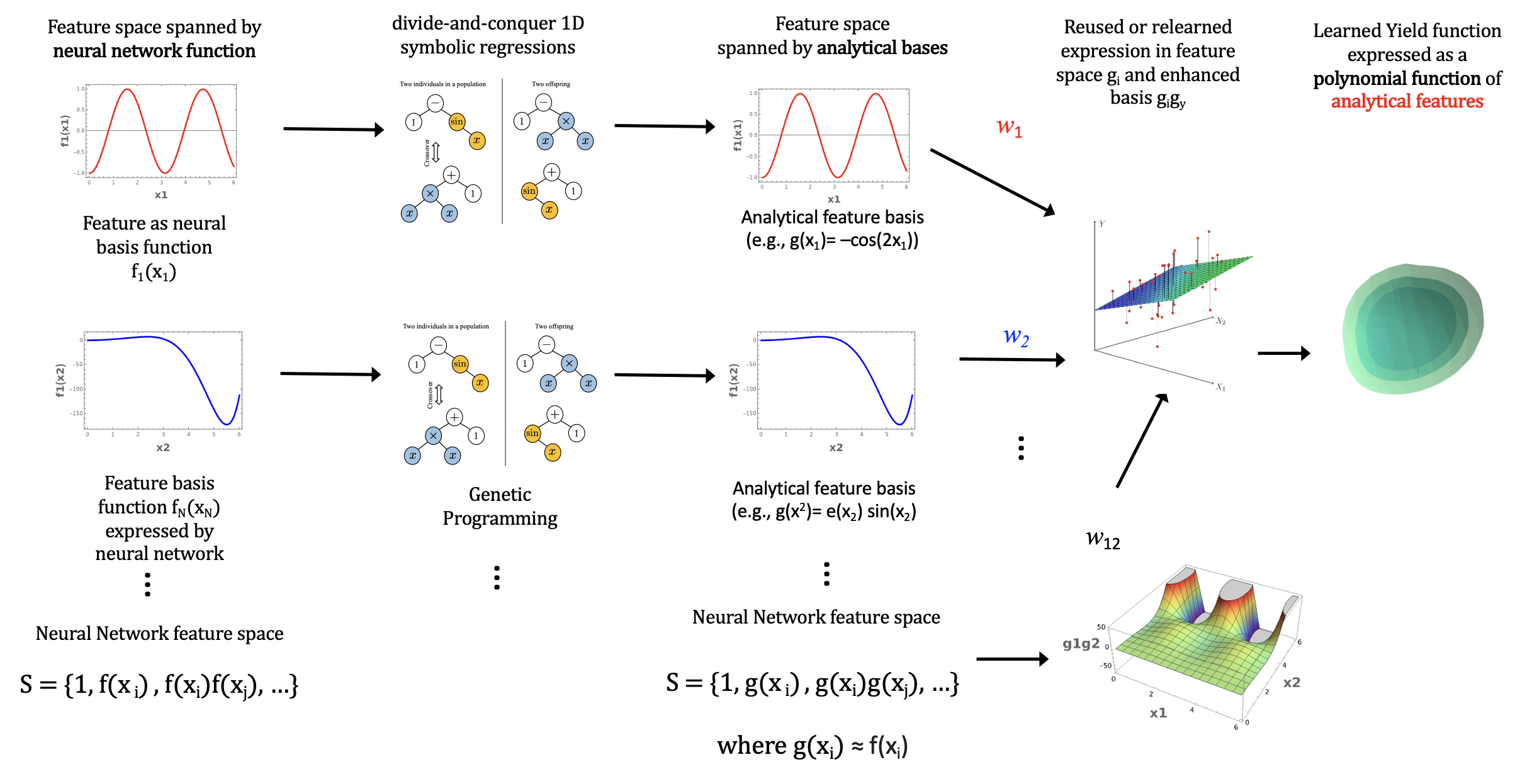}}
\caption{The divide-and-conquer symbolic regression for enhanced interpretability. A series of 1D symbolic regressions are trained to replace the 1D neural network basis function to form an analytical yield function.}
\label{fig:step2}
\end{figure}

Symbolic regression (SR) seeks to discover a mathematical expression that best fits a given data set without specifying the 
form of the mathematical expression.  Not specifying the mathematical form adds more flexibility to curve-fit the data.  However, symbolic regression, particularly for multi-dimensional vector-valued or tensor-valued functions, is significantly more difficult due to the combinatoric nature of the optimization problem necessary to search the mathematical expression \citep{icke2013improving,de2018greedy}.
However, the existence of the polynomial feature space spanned by the basis $\{1, f_{i}, f_{i} f_{j}\}$ offers us an opportunity to break down
the multi-dimensional symbolic regression problem into multiple one-dimensional problems, one for each shape function $f_{i}$.  
The final learned function is then expressed as the polynomial in the feature space $\text{span}(\{1, f_{i}, f_{i} f_{j}\})$ (see Fig. \ref{fig:step2}).
This setting may greatly reduce the difficulty of the symbolic regression problem at the expense of injecting the additional assumption that the learned function can be expressed in the aforementioned way. 

\begin{figure}[h!]
\centering
\includegraphics[height=0.35\textwidth]{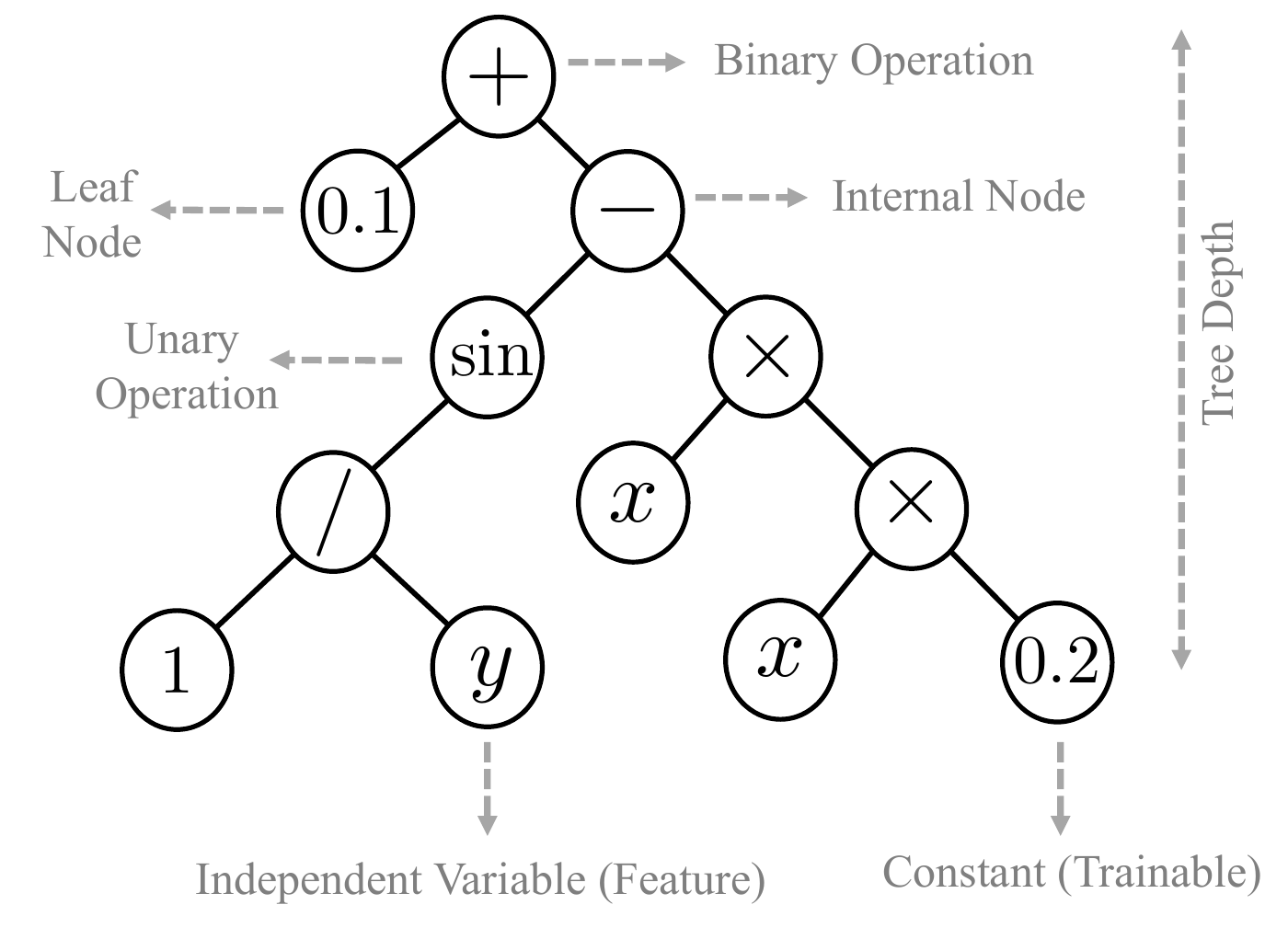}
\caption{Equation representation as expression binary tree.  This expression tree with depth 4 and size (total number of nodes) 12 represents program $\sin(\frac{1}{y}) - 0.2x^2 + 0.1$.  Expression trees are the main building blocks of modern symbolic regression algorithms.
}
\label{fig:bnary_tree}
\end{figure}

 The space of possible expressions is commonly defined by specifying the set of mathematical operators, functions, variables, and constants that can be used to construct the expressions represented efficiently in binary trees (see Fig. \ref{fig:bnary_tree}.)
Genetic programming is one of the most popular stochastic optimization methods to search the combinatorial space of all possible mathematical expressions \citep{koza1994genetic,  schmidt2009distilling,  wang2019symbolic}.
Recently, methods based on deep reinforcement learning have also been developed as alternative ways for conducting an efficient discrete search in the space of tree data structures \citep{petersen2019deep,  landajuela2021discovering}. 

In genetic programming, the space of possible expressions is represented as a \textit{population} of candidate solutions, which are randomly generated at the start of the algorithm. Each \textit{individual} candidate solution is represented as an expression binary tree (shown in Figure \ref{fig:bnary_tree}), where the leaves of the tree represent the input variables or constants, and the internal nodes represent the mathematical operations or functions. The genetic programming algorithm then evaluates the \textit{fitness} of each candidate solution by comparing its output to the target output values. Fitness measures how well the candidate solution approximates the data, and mean square error is a commonly used fitness function.

\begin{figure}[h!]
\centering
\subfigure[]
{\includegraphics[height=0.35\textwidth]{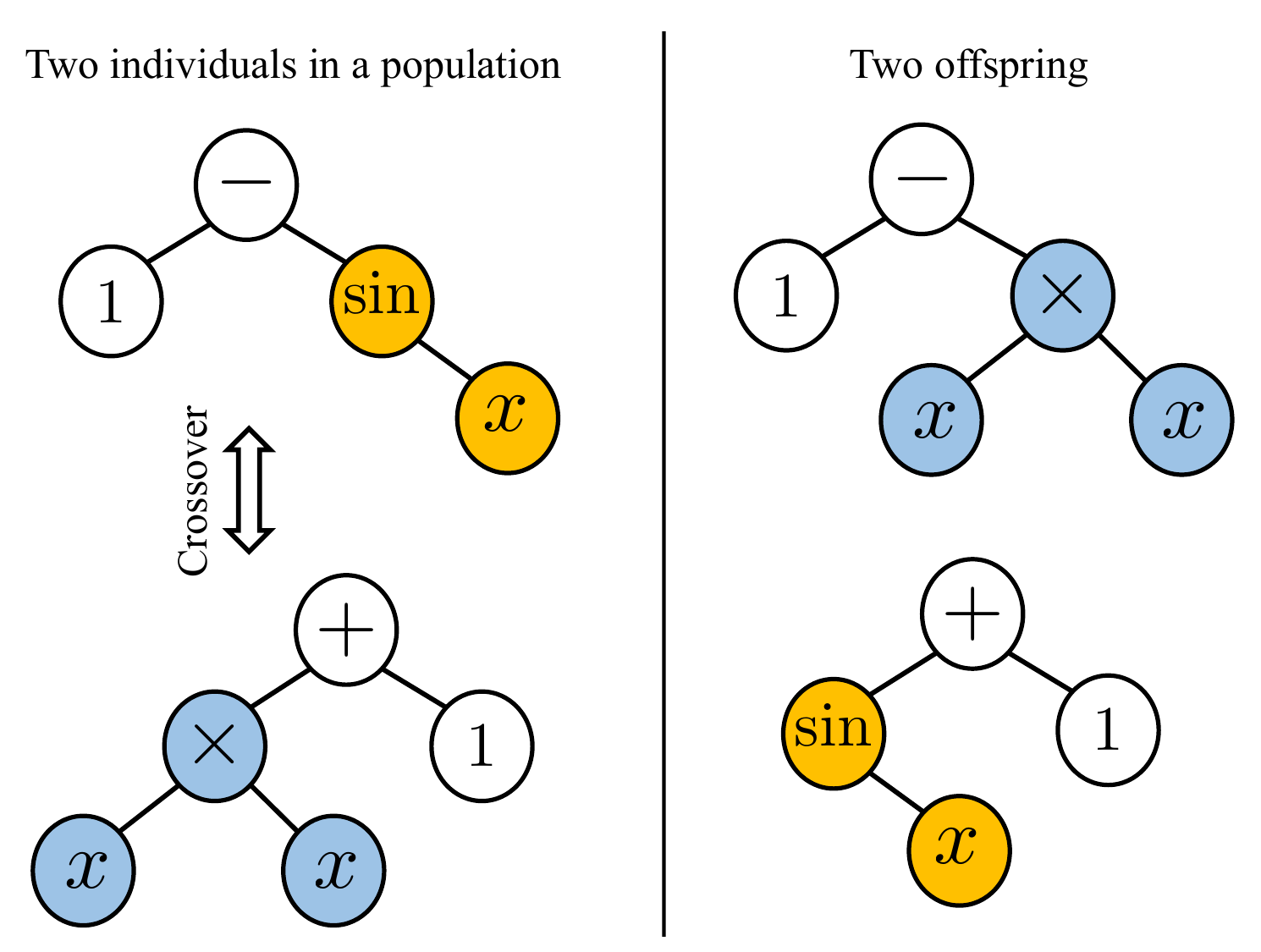}}
\subfigure[]
{\includegraphics[height=0.35\textwidth]{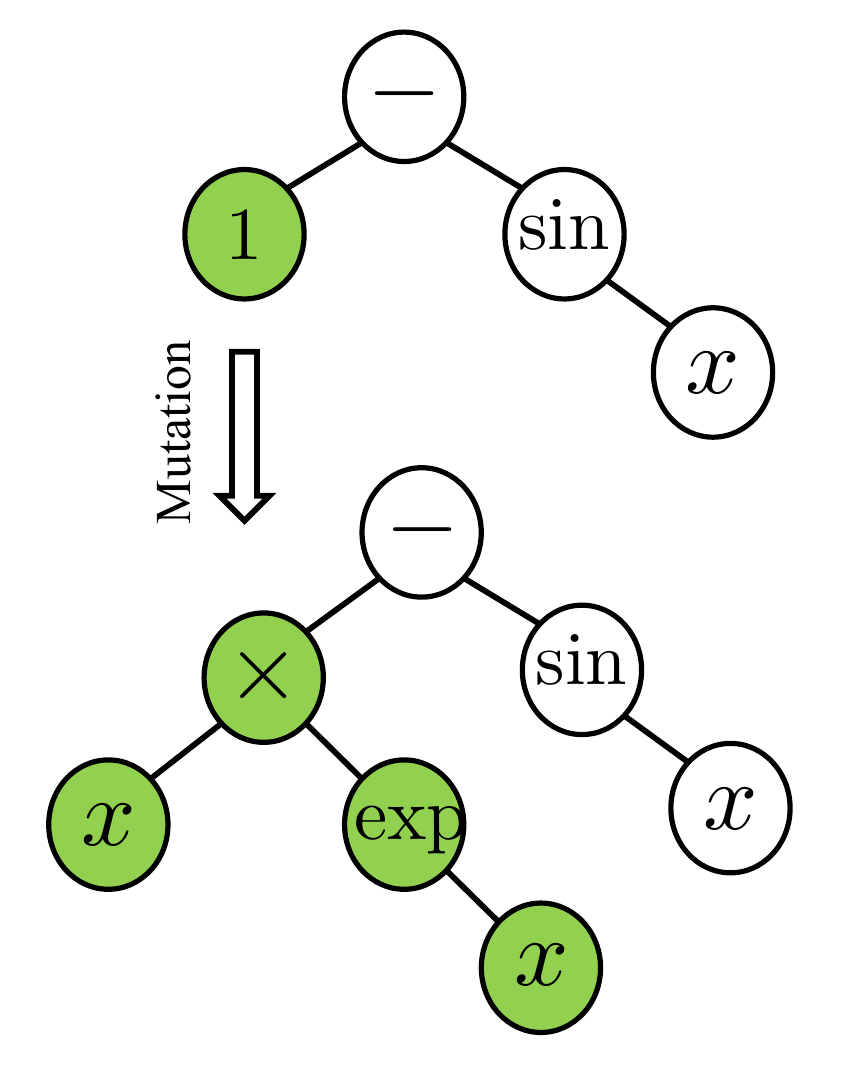}}
\caption{(a) crossover and (b) mutation operations in an evolutionary-based symbolic regression algorithm.
}
\label{fig:cross_mut_ops}
\end{figure}

The genetic programming algorithm then iteratively evolves the population of candidate solutions through \textit{selection},  \textit{crossover}, and \textit{mutation} operators in a process similar to natural selection. Selection involves choosing the fittest individuals from the current population based on their fitness scores. Crossover, as shown in Figure \ref{fig:cross_mut_ops}(a), combines the genetic information of two individuals to create offspring with characteristics from both parents.  Mutation involves randomly changing some of the genetic material of an individual to introduce new variations in the population; see Figure \ref{fig:cross_mut_ops}(b). 

Through these operations, the genetic programming algorithm creates a new generation of candidate solutions with higher fitness than the previous generation. The process is repeated until a satisfactory mathematical expression that fits the data well is found.
Once a satisfactory expression is found, it can be used to predict output values for new input values that were not used in the training dataset. At inference,  symbolic equations are more lightweight than other standard machine learning models, such as neural networks needed to store a large number of parameters, hence more transportable. Discovered equations by SR are shown to generalize well outside the train data support \citet{kim2020integration}, providing successful training.

Unlike multi-dimensional symbolic regression,  in our proposed framework,  we extract a symbolic equation for each shape function: single-variable to single-variable data.  This feature enables easy application of parallel computing; enabling SR algorithms to be executed simultaneously for each shape function.
In this work, we conduct symbolic regression with the \texttt{PySR} open-source package \citep{cranmer2020discovering,cranmer2023interpretable}, which is developed based on evolutionary-based genetic programming. In this approach, a list of unitary and binary operators can be specified to restrict the search space of tree structures. The number of adjustable constants for these operators can be specified; a gradient-based algorithm optimizes these parameters after each iteration. 
There is a tradeoff between model complexity and expressivity during SR optimization. One needs to tune and balance the complexity versus expressivity of found analytical expression to increase interpretability and reduce overfitting. Users can decide based on the desired accuracy and simplicity to choose the best equation.

\definition[Complexity score in symbolic regression]{The complexity of a symbolic equation is often assessed qualitatively rather than quantitatively, as there is no universally agreed-upon definition. In this context, we adopt the complexity measure defined in \cite{cranmer2023interpretable}, which utilizes the number of nodes in an expression tree as the complexity score. While it is possible to assign different weights to each node type, such as considering the exponential operator $\exp(\cdot)$ as more complex than the addition operator $+$, we do not incorporate such weightings in our analysis.
}

\remark[Related literature on symbolic regression for plasticity models]{
\cite{versino2017data} introduced the application of SR for learning flow stress models from data. They incorporated domain knowledge to introduce strategies such as augmenting the data and constraining the functional form to enhance the generalization accuracy of the SR.  \cite{bomarito2021development} also apply SR to discover plastic yield surface equations.
Our divide-and-conquer approach, on the other hand, did not attempt to directly infer the yield function. Instead, the focus is on applying SR to obtain expressions of a set of nonlinear univariate feature functions that form the yield function as polynomials in feature space. By limiting the SR for univariate functions, this strategy reduces the complexity of the search for optimal expression trees. In some numerical experiments, we even found that the expression could potentially be deduced manually by inspecting the patterns of the mapped feature and the input variables. 
	Additionally,  the proposed approach may exhibit superior performance in higher dimensions compared to direct SR \cite{cranmer2020discovering,petersen2019deep}; as shown in Table \ref{Tab:peterson}, the rate of equation recovery by SR reduces by increasing dimensionality. Furthermore, addressing physics constraints such as convexity can be achieved during differentiable QNM training in the first step, potentially reducing computational costs compared to the approach that incorporates physics constraints during the discrete search of SR. 
Finally,  the symbolic expression may offer a level of interpretability that makes post-hoc analysis easier to carry out (see Section \ref{appx:val-sym-convx}.)
}


\remark[Related literature on hybrid SR]{
There are other research efforts which 
 also utilized scalable models (e.g., neural networks and decision trees) in combination with SR algorithms to handle the curse of high dimensionality while maintaining interpretability \citep{icke2013improving,  cranmer2020discovering,  wadekar2020modeling, udrescu2020ai}.

\citet{cranmer2020discovering} and \citet{udrescu2020ai},  for instance,  both utilize neural networks to generate "well-motivated inductive bais" to facilitate the symbolic regression.  In the former case,  a divide-and-conquer strategy is also used to
extract conservation laws learned by graph neural networks,  incorporating a separability structure as an inductive bias in the learned function.
In the latter case, neural networks are used to find possible existing symmetries, separability, or compositional structures in the data.

\subsection{Implementation for third-party validation in open-source finite element models}
\label{sec:implications_yield_surf}

For completeness and to ensure third-party reproducibility, we outline the steps taken to implement the generated yield surface 
model into the user-defined material subroutine (UMAT) for finite element simulations.  Unlike the previous approach in \citet{suh2023publicly},  where yield functions must be parametrized via neural networks across different programming languages,  
this new implementation of the analytical model requires only the correct expression of the learned model. 
While the expression of the learned model might appear to be less elegant than the hand-crafted models,  the trade-off between simplicity 
and expressivity of the models can be adjusted,  as shown in our numerical examples in Sections \ref{sec:num_example} and \ref{sec:benchmark}.  Consequently,  the implementation of UMAT for the symbolic learned model does not require 
re-implementation of the trained neural network(s) in FORTRAN with given weights and biases, and hence much easier for any experienced production code developers and engineers.

\subsubsection{Level-set plasticity modeling framework}
\label{sec:level_set}
In the numerical examples, we limit our attention to the case where material behavior is perfectly plastic. 
As such,  the yield function $f_y$ can be expressed in terms of the Cauchy stress $\tensor{\sigma} \in \mathbb{S}$, 
while the evolution of the internal variable does not lead to the evolution of yield surface.
Nevertheless, the proposed framework may, in principle, also work for plastic hardening/softening by increasing the dimensionality of the problem to incorporate the internal variables as input variables. 
Consider the case where the yield surface is fixed in the stress space.
The elastic domain $\mathbb{E}$ is  $\{ \tensor{\sigma} \in \mathbb{S} | f_y(\tensor{\sigma}) < 0\}$, whereas the corresponding plastic domain $\partial \mathbb{E}$ is $\{\tensor{\sigma} \in \mathbb{S} | f_y(\tensor{\sigma}) = 0\}$. 
The admissible stresses belong to its closure $\overline{\mathbb{E}}$:
\begin{equation}
\label{eq:elastic_region}
\overline{\mathbb{E}} = 
\lbrace \tensor{\sigma} \in \mathbb{S} | f_y(\tensor{\sigma}) \le 0 \rbrace.
\end{equation}
As such, the stress points in the plastic regime measured either from the experiments or from sub-scale simulations always resides on the yield surface $f_y(\tensor{\sigma}) = 0$. 
Since the data lacks information inside and outside the yield surface, training a data-driven model directly from the collected stress states is not an easy task. 
For instance, the learned function may not be capable of returning positive values if the given stress is inadmissible and negative values in the elastic regime. 
Hence, this study adopts the concept of the level-set modeling framework proposed by \citet{vlassis2021sobolev,vlassis2022component} that regularizes the yield function $f_y(\tensor{\sigma})$ into a signed distance function $\phi(\tensor{\sigma})$ that is well-defined anywhere in the space of second-order symmetric tensors $\mathbb{S}$:
\begin{equation}
\label{eq:signed_distance}
\phi(\hat{\tensor{x}}) = 
\begin{dcases}
d(\hat{\tensor{x}})  &\text{if } f_y(\hat{\tensor{x}}) > 0, \\
0                    &\text{if } f_y(\hat{\tensor{x}}) = 0, \\
-d(\hat{\tensor{x}}) &\text{if } f_y(\hat{\tensor{x}}) < 0,
\end{dcases}
\end{equation}
where $\hat{\tensor{x}}$ is an arbitrary stress point represented in a proper parametric space, while $d(\hat{\tensor{x}})$ represents the minimum Euclidean distance between $\hat{\tensor{x}}$ and the yield surface in principal stress space. 
It should be noted that the choice of parametric space for representing the stress state $\hat{\tensor{x}}$ greatly affects the performance of the data-driven model, as pointed out in \citet{kuhn2013applied}. 
Although its effect will be further discussed in Section \ref{sec:low_order_NAM},  this section focuses on a cylindrical coordinate system for the $\pi$-plane orthogonal to the hydrostatic axis, i.e., $\hat{\tensor{x}} = \hat{\tensor{x}}(p, \rho, \theta)$, rather than directly adopting the Cartesian coordinates spanned by the principal stresses ($\sigma_1$, $\sigma_2$, $\sigma_3$).  In the cylindrical coordinate system, $p$ denotes the mean pressure while $\rho$ and $\theta$ are the Lode's radius and angle, respectively.

Since the yield surface cross-section perpendicular to the hydrostatic axis forms a closed loop, one possible way to construct the signed distance field $\phi$ is to solve the Eikonal equation, i.e., 
\begin{equation}
\label{eq:Eikonal}
\| \grad{\phi} \| = 1,
\end{equation}
while imposing homogeneous Dirichlet boundary condition at the stresses that belong to $\partial \mathbb{E}$. 
Based on the obtained signed distance field, we augment the original set of stress points that satisfies $f_y(\tensor{\sigma})=0$ with $N_{\phi}$ sets of points that are not necessarily located on the yield surface. 
The unit stress gradient also helps the learned model have a unit plastic flow, enabling the plastic multiplier to reflect the magnitude of the plastic strain for a given plastic flow direction. 

\subsubsection{Implicit integration of interpretable-ML-based constitutive relation}
\label{sec:stress_integration}
This section presents an implicit return mapping algorithm for the interpretable-ML-based constitutive equation that computes the stress tensor $\tensor{\sigma}_{\text{n+1}}$ at loading step $\text{n}+1$ for a given strain increment $\Delta \tensor{\varepsilon}$ and the previous stress state $\tensor{\sigma}_{\text{n}}$. 
Similar to the previous studies, e.g., \citep{wilkins1963calculation, hughes1984numerical, borja2013plasticity}, the stress integration consists of an elastic predictor that computes the trial stress $\tensor{\sigma^{\text{tr}}}_{\text{n+1}}$, followed by a plastic correction scheme, while the only difference is that we replace the mathematical expression of the yield criterion with the trained model $\bar{\phi}$ (i.e., either NAM, QNM, or symbolic model). 
In this case, by restricting the formulation within the infinitesimal range and assuming that the elasticity tensor $\mathbb{C}^e$ is given, the rate form of the constitutive equation based on an associative flow rule can be expressed as,
\begin{equation}
\label{eq:const_rate_form}
\dot{\tensor{\sigma}}
=
\mathbb{C}^e : \dot{\tensor{\varepsilon}}^e
=
\mathbb{C}^e : \left( \dot{\tensor{\varepsilon}} - \dot{\lambda} \frac{\partial \bar{\phi}}{\partial \tensor{\sigma}} \right),
\end{equation}
since the infinitesimal strain tensor can be additively decomposed into the elastic ($\tensor{\varepsilon}^e$) and plastic ($\tensor{\varepsilon}^p$) parts, while $\lambda$ indicates the plastic multiplier. 
Here, the incremental form of Eq.~\eqref{eq:const_rate_form} can be obtained by substituting the trial stress $\tensor{\sigma^{\text{tr}}}_{\text{n+1}} = \mathbb{C}^e : \tensor{\varepsilon}^{e,\text{tr}}_{\text{n+1}}$ (where $\tensor{\varepsilon}^{e,\text{tr}}_{\text{n+1}} = \tensor{\varepsilon}^e_{\text{n}} + \Delta \tensor{\varepsilon}$ indicates the trial elastic strain) computed via elastic predictor:
\begin{equation}
\label{eq:const_inc_form}
\tensor{\sigma}_{\text{n+1}}
=
\tensor{\sigma^{\text{tr}}}_{\text{n+1}}
-
\Delta \lambda \mathbb{C}^e : \left. \frac{\partial \bar{\phi}}{\partial \tensor{\sigma}} \right|_{\text{n+1}}.
\end{equation}
If we further limit our attention to the case where the plastic behavior of our target material is isotropic, the predictor-corrector scheme can be reduced in principal stress axes as:
\begin{equation}
\label{eq:const_principal_axes}
\sigma_A = \sigma_A^{\text{tr}} - \Delta \lambda \sum_{B=1}^3 C^e_{AB} \frac{\partial \bar{\phi}}{\partial \sigma_A}
\: \: ; \: \:
\varepsilon^e_A = \varepsilon^{e,\text{tr}}_A - \Delta \lambda \frac{\partial \bar{\phi}}{\partial \sigma_A},
\end{equation}
where we omit the subscript $\text{n}+1$ for brevity. 
Here, $\sigma_A$ ($A = \lbrace 1,2,3 \rbrace$) denotes the principal stress, and $C^e_{AB}$ indicates the elastic moduli in principal axes, where its matrix form can be expressed as,
\begin{equation}
\label{eq:matrix_elas_mod}
[C_{AB}^e]
=
\begin{bmatrix}
K + \frac{4 \mu}{3} & K - \frac{2 \mu}{3} & K - \frac{2 \mu}{3} \\[1.2ex]
K - \frac{2 \mu}{3} & K + \frac{4 \mu}{3} & K - \frac{2 \mu}{3} \\[1.2ex]
K - \frac{2 \mu}{3} & K - \frac{2 \mu}{3} & K + \frac{4 \mu}{3} 
\end{bmatrix},
\end{equation}
if the elastic behavior of the material is linear, while $K$ and $\mu$ are the bulk and shear moduli, respectively. 
Recall that either NAM or its polynomial extension (QNM) adopts a total of $D$ univariate MLPs assigned for each input feature, while we parametrize stresses in cylindrical coordinates, e.g., $\bar{\phi} = \bar{\phi}(\hat{\tensor{x}})$. 
Hence, the stress gradient of the trained model in Eq.~\eqref{eq:const_principal_axes} can be obtained via the chain rule as,
\begin{equation}
\label{eq:stress_grad1}
\frac{\partial \bar{\phi}}{\partial \sigma_A}
=
\sum_{i=1}^3
\frac{\partial \bar{\phi}}{\partial \hat{x}_i} \frac{\partial \hat{x}_i}{\partial \sigma_A},
\end{equation}
where:
\begin{equation}
\label{eq:stress_grad2}
\frac{\partial \bar{\phi}}{\partial \hat{x}_i} 
= 
\left[ w_i + \underbrace{w_{ii} f_i(\hat{x}_i) + \sum_{j=1}^D w_{ij} f_j(\hat{x}_j)}_{\text{H.O.T.}} \right] 
\frac{\partial f_i}{\partial \hat{x}_i}
\: \: \text{(no sum)}.
\end{equation}
Note that higher order terms (H.O.T.) in Eq.~\eqref{eq:stress_grad2} only exist if we adopt the QNM-based symbolic model. 
Based on Eq.~\eqref{eq:const_principal_axes} and the consistency condition $\bar{\phi} = 0$, we formulate a return mapping algorithm in the principal strain space which iteratively solves a nonlinear problem $\tilde{\vec{r}}(\tilde{\vec{x}}) = \vec{0}$ until the magnitude of the residual vector reaches an acceptable value near zero. 
Specifically, we construct the local residual vector $\tilde{\vec{r}}(\tilde{\vec{x}})$ and the unknown vector $\tilde{\vec{x}}$ as follows:
\begin{equation}
\label{eq:residual_and_solution_vec}
\tilde{\vec{r}}(\tilde{\vec{x}})
=
\begin{bmatrix}
\tensor{\varepsilon}_1^e - \tensor{\varepsilon}_1^{e,\text{tr}} + \Delta \lambda \frac{\partial \bar{\phi}}{\partial \sigma_1} \\[1.2ex]
\tensor{\varepsilon}_2^e - \tensor{\varepsilon}_2^{e,\text{tr}} + \Delta \lambda \frac{\partial \bar{\phi}}{\partial \sigma_2} \\[1.2ex]
\tensor{\varepsilon}_3^e - \tensor{\varepsilon}_3^{e,\text{tr}} + \Delta \lambda \frac{\partial \bar{\phi}}{\partial \sigma_3} \\[1.2ex]
\bar{\phi}(\hat{\tensor{x}})
\end{bmatrix}
\: \: ; \: \:
\tilde{\vec{x}}
=
\begin{bmatrix}
\tensor{\varepsilon}_1^e \\
\tensor{\varepsilon}_2^e \\
\tensor{\varepsilon}_3^e \\
\Delta \lambda
\end{bmatrix},
\end{equation} 
such that the admissible Cauchy stress tensor at loading step $\text{n}+1$ can be recovered once we obtain the converged set of solutions $\tilde{\vec{x}}$, e.g.,
\begin{equation}
\label{eq:stress_recov}
\tensor{\sigma}_{\text{n+1}}
= 
\mathbb{C}^e : \left[ \sum_{A=1}^{3} \varepsilon^e_A (\vec{n}_A \otimes \vec{n}_A) \right],
\end{equation}
where $\vec{n}_A$ ($A = \lbrace 1,2,3 \rbrace$) indicates the principal direction.

\section{Results}
\label{sec:num_example}
%
%
In this section, we use three representative numerical examples to demonstrate the feasibility of  the proposed interpretable framework for the data-driven discovery of yield surfaces as well as benchmark the performance of the discovered models both at material point and PDE simulations. 
The first example in Section \ref{sec:low_order_NAM} focuses on a pressure-insensitive dataset and examines how the proposed method can discover the symbolic yield surface. We also examine the method's extrapolation capability compared to previous methods in the literature. 
In the second example,  Section \ref{sec:high_order_NAM}, we study the efficacy of the QNM-based symbolic regression method in dealing with higher-dimensional data of metal plasticity in a five-dimensional space. We also discuss the enforcement of simplicity through sparsity control. The last example in Section \ref{sec:FE_simulation} illustrates how the proposed QNM can discover an accurate symbolic equation for pressure-sensitive data. We demonstrate that the found symbolic equations can be readily used in classical FEM codes without significant changes compared to neural network-based plasticity models.

\subsection{isochoric elastoplasticity models for benchmark performance}
\label{sec:low_order_NAM}
In this example, we use pressure-insensitive data in the stress space, which has three dimensions. However, due to pressure insensitivity, only two dimensions are required to describe the yield surface. Our modeling framework aims to determine whether it can distinguish this independence. We also investigate the impact of spectral layer and data parameterization on training performance. Furthermore, we demonstrate that correct assumptions and inductive biases can improve generalization by conducting stress point integration via the return mapping algorithm.  

The benchmark function where we generate a set of synthetic stress points resembles the von Mises yield criterion. 
While it manifests 
a cylinder shape along the hydrostatic axis, the benchmark yield surface is also dependent on the Lode's angle $\theta$ such that it exhibits a flower-shaped cross-section: 
\begin{equation}
\label{eq:benchmark_flower_shape}
f = 
\sqrt{\frac{3}{2}} \rho
\left[
1 + A_p \sin(k_p \theta)
\right] - \sigma_y,
\end{equation}
where the parameters $k_p$ and $A_p$ control the number and the size of petals, respectively, while $\sigma_y$ is the yielding stress. 
From Eq.~\eqref{eq:benchmark_flower_shape}, we choose the parameters as $k_p = 3$, $A_p = 0.325$, and $\sigma_y = 250$ MPa, and then collect a set of stress points that satisfies $f = 0$. 
Specifically, we sample 20 data points along the mean pressure axis (from $-1$ GPa to $1$ GPa) and 120 points along the Lode's angle axis (from $0$ to $2 \pi$), such that a total of 2,400 different admissible stress states are considered as an original dataset. 
Then the original data points are then pre-processed via the signed distance function by setting $N_{\phi} = 11$, such that the number of stress points in our full dataset is 26,400. 
Here, compared to the previous studies \citep{vlassis2021sobolev, vlassis2022component} where the Lode's radii of the training dataset vary from 0 to $2 \rho$, as illustrated in Figure \ref{fig:flower_level_set}, our augmented dataset only covers a narrow band region for the original yield surface (i.e., [0.85$\rho$, 1.15 $\rho$]) in order to test the extrapolation capability of our symbolic regression model obtained from the trained NAM. 

\begin{figure}[h]
\centering
\includegraphics[height=0.4\textwidth]{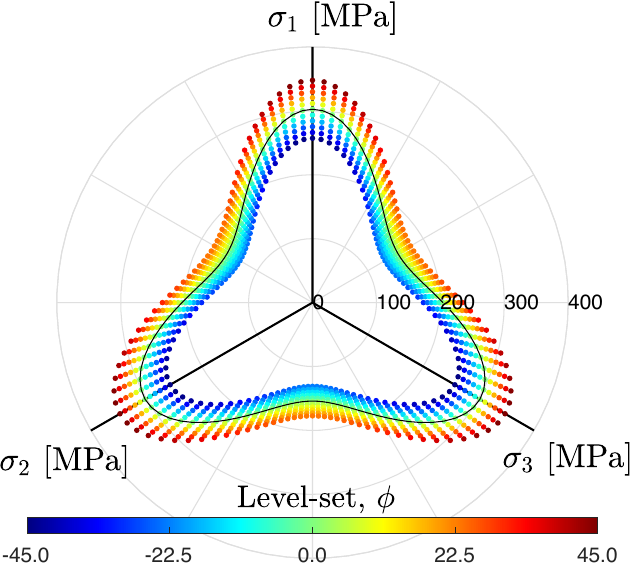}
\caption{Training dataset (colored symbols) augmented from the original dataset that satisfies $f = 0$ (black curve).}
\label{fig:flower_level_set}
\end{figure}

Our experiments suggest that the NAM setting is sufficient to recover the correct yield surface in this problem. Therefore,  we focus on presenting the NAM results and omit the QNM results for brevity.  In Fig.\ref{fig:loss_spec_nspec_full_data}, we compare two NAM models trained with the same number of parameters (one with a Fourier layer and one without) as the control experiment. Both models are trained with the full data set. The results suggest that the network with the Fourier layer achieves higher accuracy with fewer iterations. 

\subsubsection{Performance with small dataset}
To avoid false positives in our findings, we repeat the experiments, but this time trained the two models with a randomized dataset.
This random dataset is a subset of the entire dataset,  which consists of only 2,000 points on the yield surface and 3,000 points that yield non-zero level-set values.  In this second case where data is sparser (see the results in Fig.\ref{fig:loss_spec_nspec_less_data}. ),  we observe a significantly greater  performance difference between the two methods. 

\begin{figure}[h]
\centering
\subfigure[]{\label{fig:loss_spec_nspec_full_data}
\includegraphics[height=0.375\textwidth]{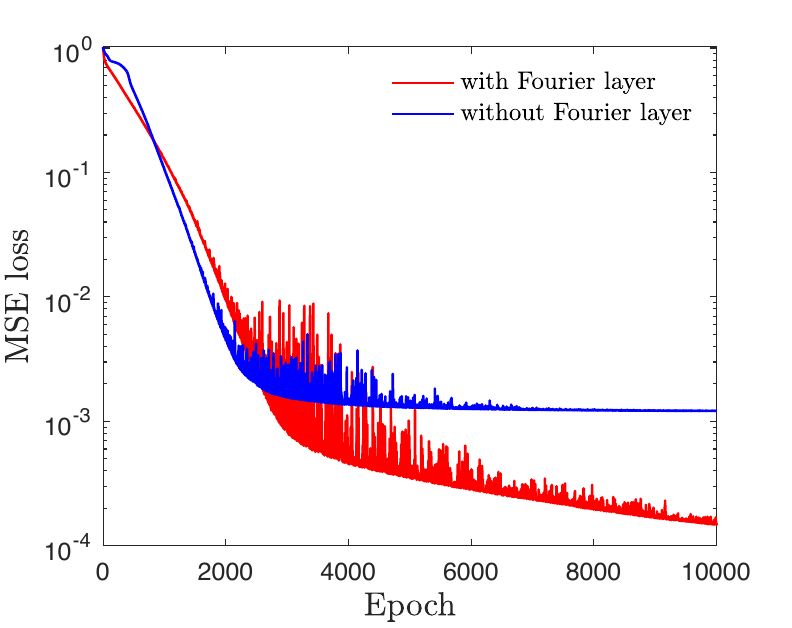}}
\hspace{0.01\textwidth}
\subfigure[]{\label{fig:loss_spec_nspec_less_data}
\includegraphics[height=0.375\textwidth]{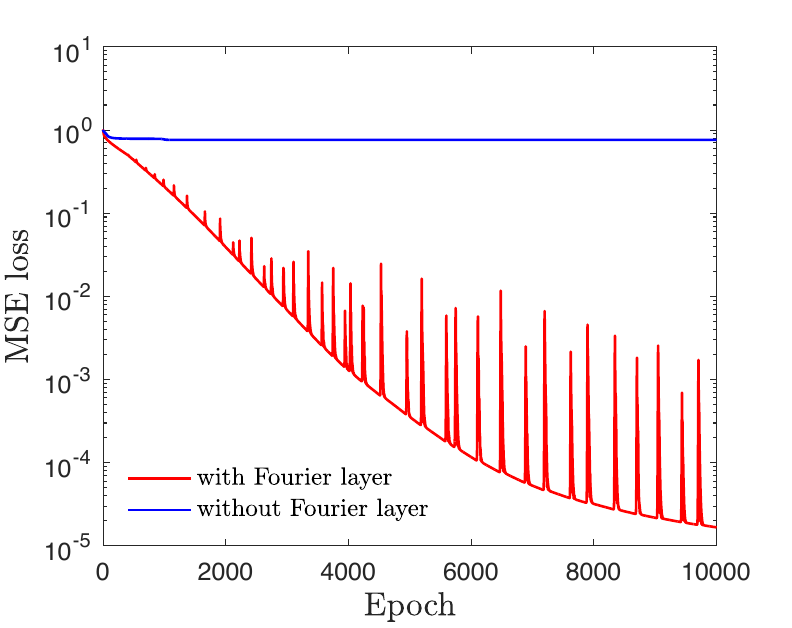}}
\caption{Comparing training performance between neural networks used the Fourier layer v.s. not used: (a) using all the data; (b) using fewer data.}
\label{fig:loss_spec_nspec}
\end{figure}

In Fig.~\ref{fig:loss_parameterization}, we study the effect of input data representation on the learning task using the same neural network architecture for both cylindrical and Cartesian coordinate systems, with the spectral layer utilized. The results suggest that the cylindrical coordinate system can outperform the Cartesian coordinate system, making a difficult training process much easier. Finding an appropriate data representation may become more critical in our proposed framework based on the NAM or QNM, as we have stronger assumptions regarding feature separability compared to classical surrogate modeling methods; our approach is also less flexible than those methods. This is consistent with classical approaches in mechanics, where researchers have introduced different coordinate systems to transform a complex problem into an easier one in the new coordinate system, for example, by taking into account the underlying symmetries in the new coordinate system.

\begin{figure}[h]
\centering
\includegraphics[height=0.375\textwidth]{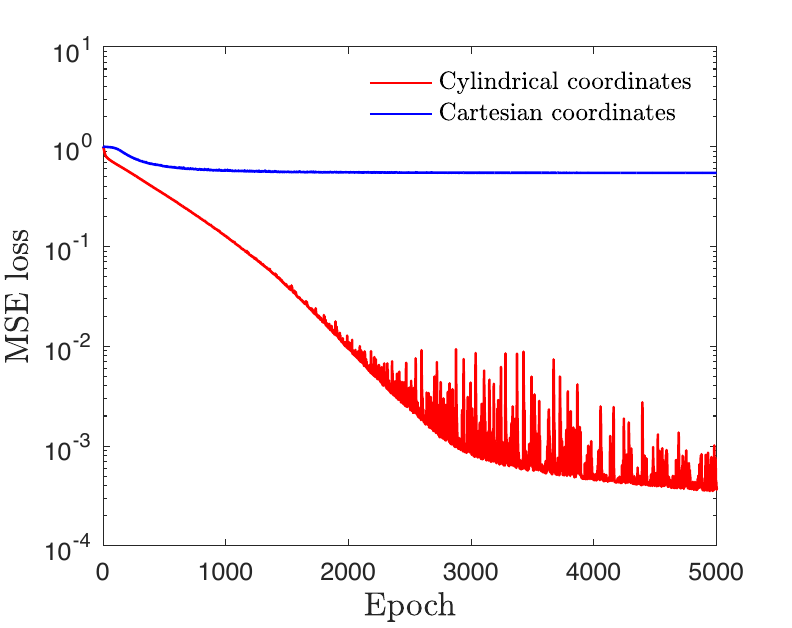}
\caption{Prediction loss v.s.  training epochs when stress data is represented in cylindrical and Cartesian coordinate systems.}
\label{fig:loss_parameterization}
\end{figure}

We focus solely on the model trained with the full data set in the cylindrical coordinate system that utilized the Fourier layer. Figure~\ref{fig:flower_shape_funcs} displays the learned shape functions by NAM after training their associated symbolic equations extracted by the symbolic regression algorithm. An advantageous feature of the NAM or QNM modeling idea is its ability to find appropriate univariate, separable representations of complex, multivariate data. This allows each shape function to be visually inspected individually, and even a reasonable symbolic equation can be derived for each univariate data set. In this example, it is clear by inspection that constant, linear, and sinusoidal functions can describe the NAM shape functions reasonably well. This is one of the primary advantages of using a divide-and-conquer algorithm to break down complexities into more straightforward tasks that can be handled more efficiently by humans.  

The weights associated with each shape function are as follows: $w_p = 0.43$, $w_{\rho} = 5.27$, and $w_{\theta}=3.82$, where these weights are denoted as $w_i$ in Eq.\ref{eq::nam}. Notably, the weight corresponding to the pressure coordinate is about one order of magnitude less than the weights of the other shape functions. Furthermore, as seen in Fig.\ref{fig:flower_shape_funcs}(a), the pressure shape function behaves almost constantly around 1. These observations confirm that the NAM is capable of discarding the effect of pressure coordinate on the final prediction, which is expected since the data is pressure-insensitive.

\begin{figure}[h]
\centering
\subfigure[]{\label{fig:flower_fp}\includegraphics[height=0.255\textwidth]{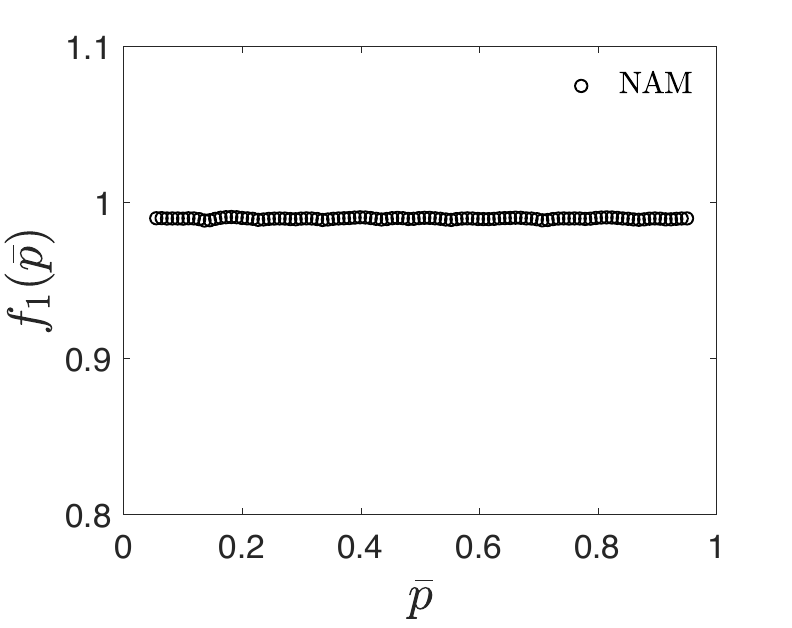}}
\hspace{0.01\textwidth}
\subfigure[]{\label{fig:flower_frho}\includegraphics[height=0.255\textwidth]{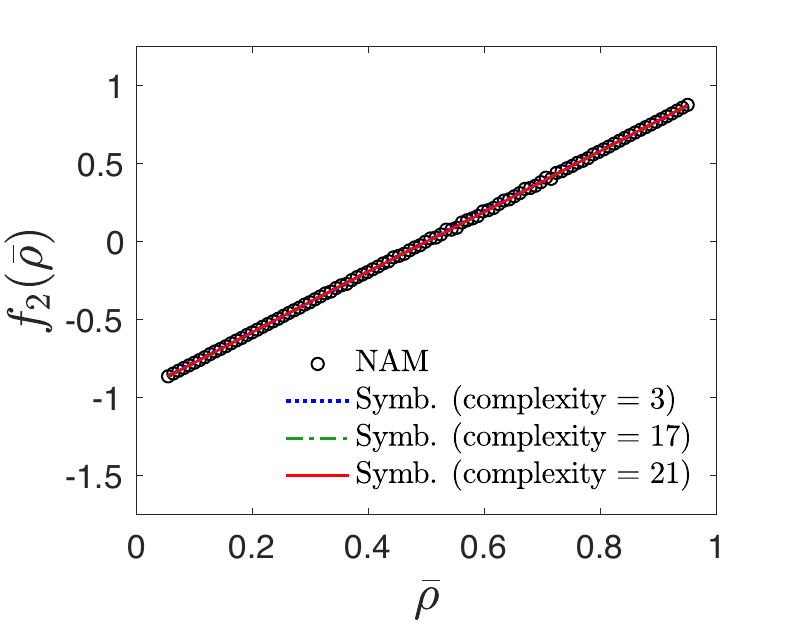}}
\subfigure[]{\label{fig:flower_ftheta}\includegraphics[height=0.255\textwidth]{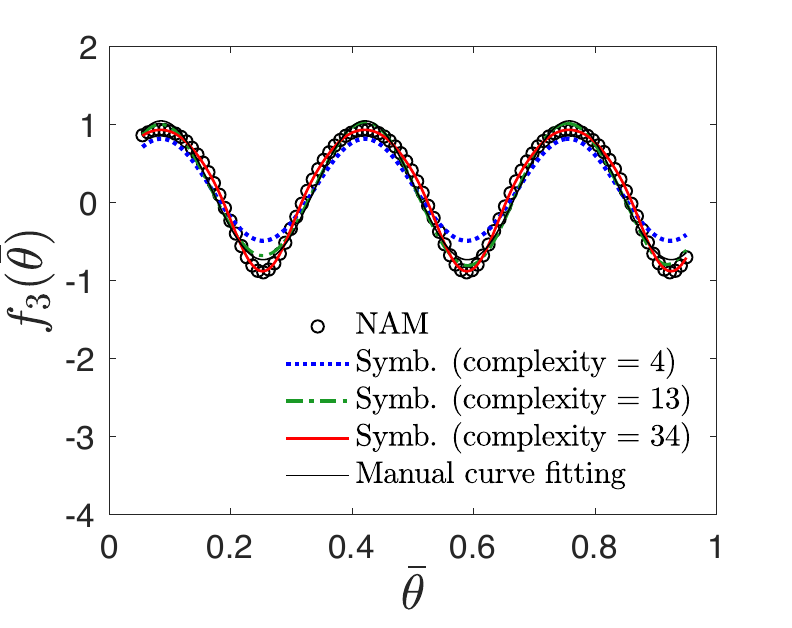}}
\caption{Shape functions learned by neural network and extracted accordingly by the symbolic regression algorithm: (a) the function corresponding to the normalized pressure $\bar{p}$,  (b) the function corresponding to the normalized radius $\bar{\rho}$, (c) the shape function corresponding to the normalized angle $\bar{\theta}$.  Normalization in this study is a linear transformation of data into the range $[0, 1]$.  Complexity labels indicate the level of complexity for the selected symbolic equations.  Higher complexity is an indication of more terms.  These symbolic equations are shown in Tables~ \ref{Tab::petal-symb-eqs-rho} and \ref{Tab::petal-symb-eqs-theta}.}
\label{fig:flower_shape_funcs}
\end{figure}

We apply the symbolic regression algorithm to determine the remaining shape functions, allowing for flexibility in equation forms. The algorithm employs binary and unary operations, including addition, multiplication, division, cosine, exponential, sine, and logarithm. We selected equations with varying complexities and displayed them in Figs.\ref{fig:flower_shape_funcs}(b,c), along with their explicit forms listed in Tables \ref{Tab::petal-symb-eqs-rho} and \ref{Tab::petal-symb-eqs-theta}.

The second shape function $f_2(\bar{\rho})$ exhibits almost the same accuracy as the least complex equation, which is a linear function. It is noteworthy that the other options in Table \ref{Tab::petal-symb-eqs-rho} with higher complexity scores include a linear term and a sinusoidal function. However, the amplitude of the sinusoidal term is two orders of magnitude smaller than that of the linear term, allowing it to be ignored.
The optimal trade-off between accuracy and simplicity for the third shape function $f_3(\bar{\theta})$ in Table \ref{Tab::petal-symb-eqs-theta} is less apparent, as learned function of higher complexity is significantly more accurate.

In certain cases, one may rely on intuition to identify the appropriate equation without using the symbolic regression algorithm. This advantage stems from the one-dimensional nature of curve-fitting tasks. For instance, one may hypothesize that the third shape function is a sinusoidal function of the form $f_3(\bar{\theta}) = a \sin( b \pi \bar{\theta}) + c) + d$ and determine the unknown parameters $a$, $b$, $c$, and $d$ through a nonlinear least squares method. In this case, we obtain $f_3(\bar{\theta})= 0.89 \sin(5.95\pi \bar{\theta}- 0.02) + 0.16$, which is labeled "manual curve fitting" in Fig. \ref{fig:flower_shape_funcs}(c). In terms of the trade-off between complexity and accuracy, one may prefer this equation over those obtained through symbolic regression algorithms, which are, in fact, closer to the ground truth function, Eq.~\eqref{eq:benchmark_flower_shape}. This simple exercise demonstrates that human intuition may outperform symbolic regression algorithms.

\begin{table}
  \centering
  \caption{Found symbolic shape function $f_2(\bar{\rho})$ for pressure-insensitive benchmark}\vspace{-10pt}
        \begin{tabular}{|p{10cm}|c|c|}
        \hline
        Expression & Complexity score & Loss\\[3mm]
          \hline
        & & \\
        $\begin{aligned}f_2(\bar{\rho}) = 1.0 \bar{\rho}\end{aligned}$ & 3 & 5.546e-05\\[5mm] 
        $\begin{aligned}
        f_2(\bar{\rho}) = \bar{\rho} - 0.01 \sin{\left(\sin{\left(\sin{\left(\bar{\rho} + \sin{\left(\bar{\rho} \right)} \right)} \right)} \right)} \cos{\left(1.32 \bar{\rho} \right)}\end{aligned}$ & 17 & 4.908e-05 \\[5mm]
        $\begin{aligned}
        f_2(\bar{\rho}) =\bar{\rho} - 0.01 \sin{\left(\sin{\left(0.77 \bar{\rho} + \sin{\left(\sin{\left(\bar{\rho} \right)} \right)} + 0.29 \right)} \right)} \cos{\left(1.32 \bar{\rho} \right)}\end{aligned}$ & 21 & 4.894e-05 \\ \hline\end{tabular}
\label{Tab::petal-symb-eqs-rho} 
\end{table}

\begin{table}
  \centering
  \caption{symbolic shape function $f_3(\bar{\theta})$ for pressure-insensitive benchmark}\vspace{-10pt}
        \begin{tabular}{|p{10cm}|c|c|}
        \hline
        Expression & Complexity score & Loss \\[3mm]
        \hline
      & &   \\
        $\begin{aligned}f_3(\bar{\theta}) = - \sin{\left(4.84 \bar{\theta} \right)}\end{aligned}$& 4 & 9.466e-02\\[5mm] 
        $\begin{aligned}
        f_3(\bar{\theta}) =- \frac{\sin{\left(4.83 \bar{\theta} \right)}}{\cos{\left(\sin{\left(\cos{\left(\sin{\left(\sin{\left(\cos{\left(e^{\bar{\theta}} \right)} \right)} \right)} \right)} \right)} \right)}}\end{aligned}$ & 13 & 2.367e-02\\[10mm]  
        $\begin{aligned}
        &f_3(\bar{\theta}) =
        -1.36 \sin{\left(4.83 \bar{\theta} \right)} + 1.36 \cos(\left(0.86 \sin{\left(4.83 \bar{\theta} \right)} \right.\\ 
        &\left. - 0.86 \cos{\left(0.81 \sin{\left(\sin{\left(4.83 \bar{\theta} \right)} \right)} \right)} + 0.69 \right)) - 1.1\end{aligned}$ & 34 & 3.094e-04 \\[5mm]  \hline\end{tabular}
\label{Tab::petal-symb-eqs-theta} 
\end{table}

Figure \ref{fig:yield_surf_flower} illustrates that our symbolic regression model (red curve) is capable of reproducing the shape of the benchmark yield function (black curve). 
Here, based on the full dataset, we also train a single multivariate MLP that consists of a number of fully connected layers and Multiply layers \citep{vlassis2021sobolev} to compare the predictive capability against our proposed framework. 
Although a multivariate MLP trained based upon the level-set augmented data can capture the yield surface (blue dots) that is similar to the benchmark, however, it fails to reproduce the stress-strain curve based upon an implicit stress integration scheme [Figure \ref{fig:stress_strain_flower}] due to its limited capacity to make predictions outside the training domain [0.85$\rho$, 1.15$\rho$]. 
On the other hand, as illustrated in Figure \ref{fig:stress_strain_flower}, the symbolic regression model results in a stress-strain curve that is nearly identical to the benchmark (except the approximation of $\pi$ value).  This result suggests that the algorithm is capable of finding an expression that makes accurate predictions outside the range of the training data based upon its extrapolation capacity. 

\begin{figure}[h]
\centering
\subfigure[]{\label{fig:yield_surf_flower}
\includegraphics[height=0.375\textwidth]{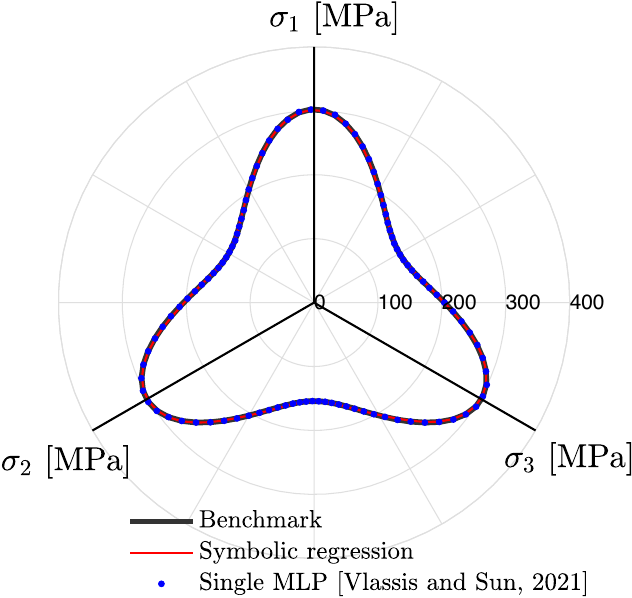}}
\hspace{0.01\textwidth}
\subfigure[]{\label{fig:stress_strain_flower}
\includegraphics[height=0.375\textwidth]{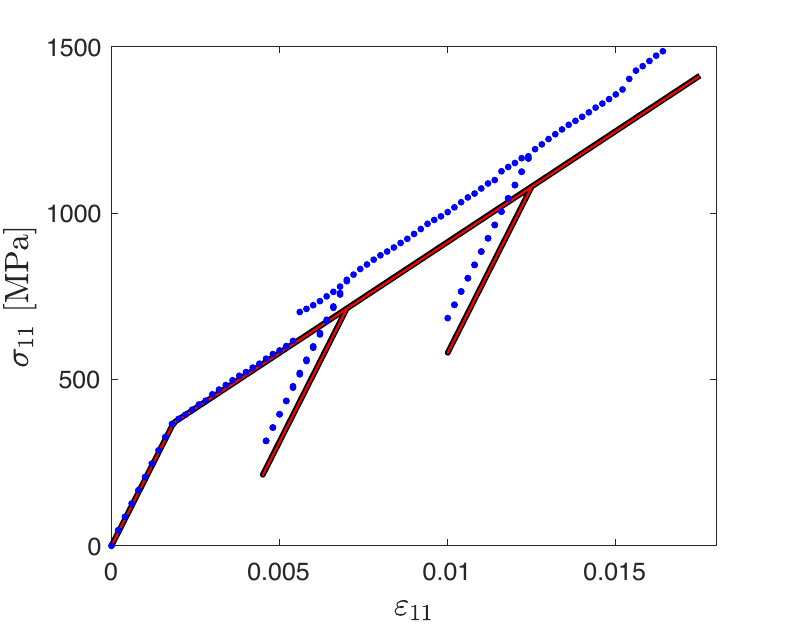}}
\caption{(a) the recovered yield surface by the introduced scheme and single multi-input MLP; (b) the stress-strain curve obtained by the return mapping algorithm for the ML-based yield surfaces.  The discovered symbolic yield surface offers a good extrapolation capability for loading conditions beyond the range of the training data in comparison to the purely neural network-based yield surface.
}
\label{fig:flower_comparison}
\end{figure}

\subsection{Discovery of symbolic level set plasticity model from noisy porous metal data}
\label{sec:high_order_NAM}
In this section, we benchmark the application of the proposed method for finding the plastic yield surface of porous metal material. The data in this problem are in five-dimensional space, including the level-set. We will discuss equation discovery under sparsity control.  

In this section, we chose a model that was discovered by \citet{bomarito2021development}, which describes the plastic behavior of a porous material depending on the hydrostatic pressure $\bar{\sigma}_h$, the von Mises stress $\bar{\sigma}_{vm}$, the volume-averaged Lode parameter $\bar{L} = 3 \sqrt{3} (\sigma_1 - \bar{\sigma}_h) (\sigma_2 - \bar{\sigma}_h) (\sigma_3 - \bar{\sigma}_h) / (2 J_2^{3/2})$, and a parameter $\bar{v}$ that describes the void fraction. 
To generate the training data, we adopt the expression for the yield function that can be found in Eq. (48) in \citep{bomarito2021development}, and sample 20 data points along the $p$-axis (from 0 to 1.8 MPa), 30 points along the $\theta$-axis (from 0 to 2$\pi$) based on the cylindrical coordinate system, and 10 points along the $\bar{v}$-axis (from 0.063 to 0.065) such that in total 6,000 different admissible stress states are considered. 

Here,  we add uniformly distributed noise along the radial direction where its magnitude ranges from $-4$ \% to $4$ \% of Lode's radius to test the performance of the model trained by a dataset with noise.  Similar to the previous example,  the original dataset is then pre-processed via level-set augmentation by setting $N_{\phi} = 11$ that covers a narrow band region of [0.85$\rho$, 1.15$\rho$] such that our full dataset consists of 66,000 stress points. 

Figure~\ref{fig:bomarito_lo_vs_ho} displays the yield surfaces found at different levels of hydrostatic stress and void volume fraction using the NAM and QNM methods.  Both methods generate models that accurately represent the underlying yield surfaces, but the QNM method performs slightly better, especially at higher levels of hydrostatic stress, due to its higher level of flexibility. The QNM results shown in this figure were obtained by training the model with sparsity control, with $\alpha_{\text{lo}} = 0.01$ and $\alpha_{\text{ho}} = 0.001$, see Eq.~\ref{eq::sparse_loss}. The QNM method uses four learnable shape functions, each with an associated learnable weight ($w_1$, $w_2$, $w_3$, and $w_4$). The complete quadratic approximation based on these four shape functions has ten additional terms that are controlled by trainable weights ($w_{ij}$) for $1\le i \le j \le 4$.

In Figure~\ref{fig:sparsity_control}, we can see how the weights change during training when sparsity is enforced compared to when it is not. When sparsity control is used, all of the lower-order contributions (shown by different colors) eventually diminish, with $w_i$ approaching zero in later epochs, improving the model's simplicity and interpretability. Notably, this is consistent with the benchmark equation, which features couplings between multiple features and does not involve any single-variable term.

\begin{figure}[h]
\centering
\subfigure[]{\label{fig:bomarito_lo_vs_ho1}
\includegraphics[height=0.375\textwidth]{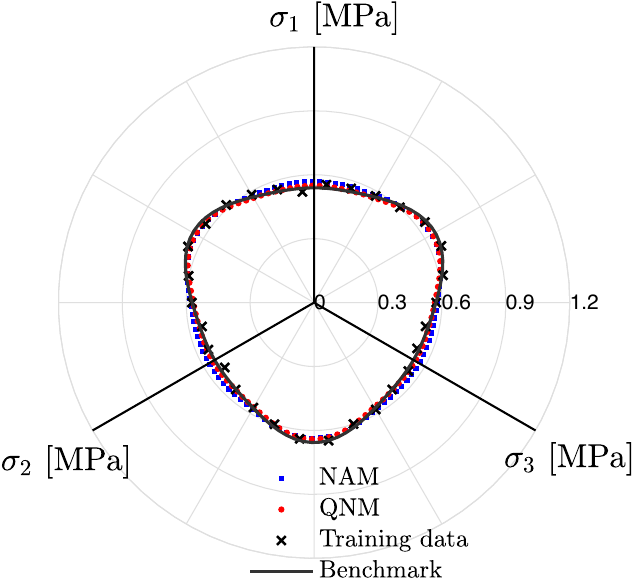}}
\hspace{0.01\textwidth}
\subfigure[]{\label{fig:bomarito_lo_vs_ho2}
\includegraphics[height=0.375\textwidth]{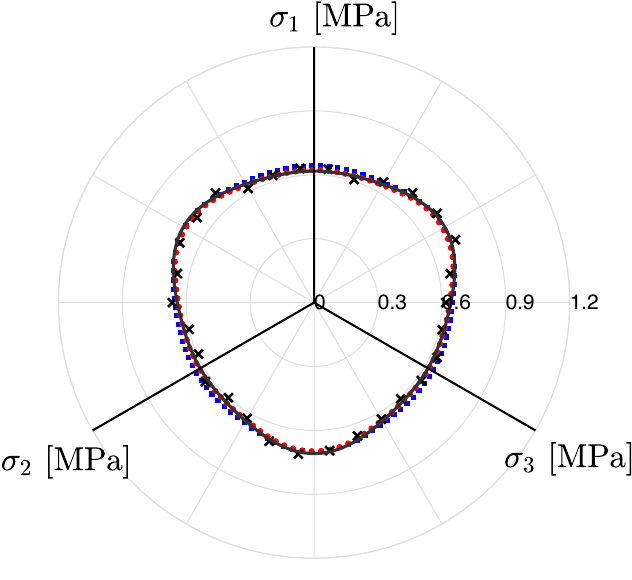}}
\vspace{0.01\textwidth}
\subfigure[]{\label{fig:bomarito_lo_vs_ho3}
\includegraphics[height=0.375\textwidth]{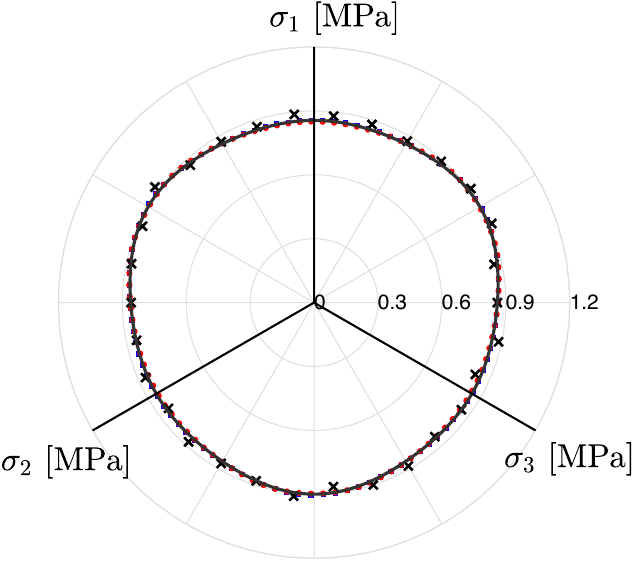}}
\hspace{0.01\textwidth}
\subfigure[]{\label{fig:bomarito_lo_vs_ho4}
\includegraphics[height=0.375\textwidth]{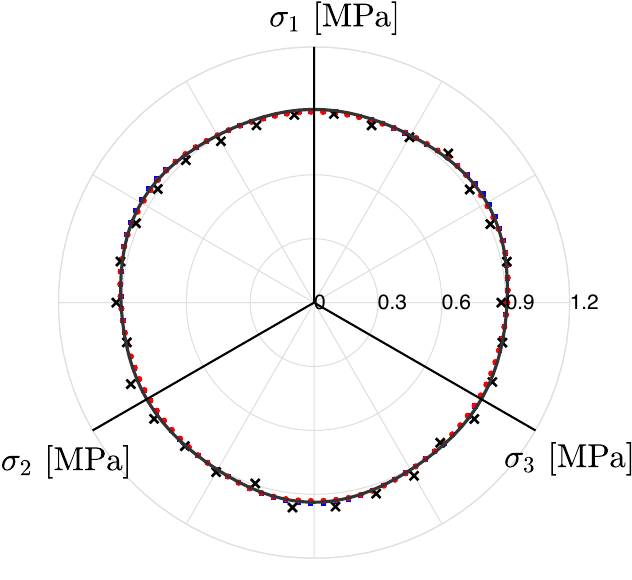}}
\caption{Comparison between NAM and QNM: (a) $\bar{\sigma}_h = 1.4210$ MPa, $\bar{v} = 0.0638$; (b) $\bar{\sigma}_h = 1.2315$ MPa, $\bar{v} = 0.0645$; (c) $\bar{\sigma}_h = 0.6631$ MPa, $\bar{v} = 0.0636$; (d) $\bar{\sigma}_h = 0.4736$ MPa, $\bar{v} = 0.0641$.}
\label{fig:bomarito_lo_vs_ho}
\end{figure}

\begin{figure}[h]
\centering
\subfigure[]{\label{fig:weight_w_sparse}
\includegraphics[height=0.375\textwidth]{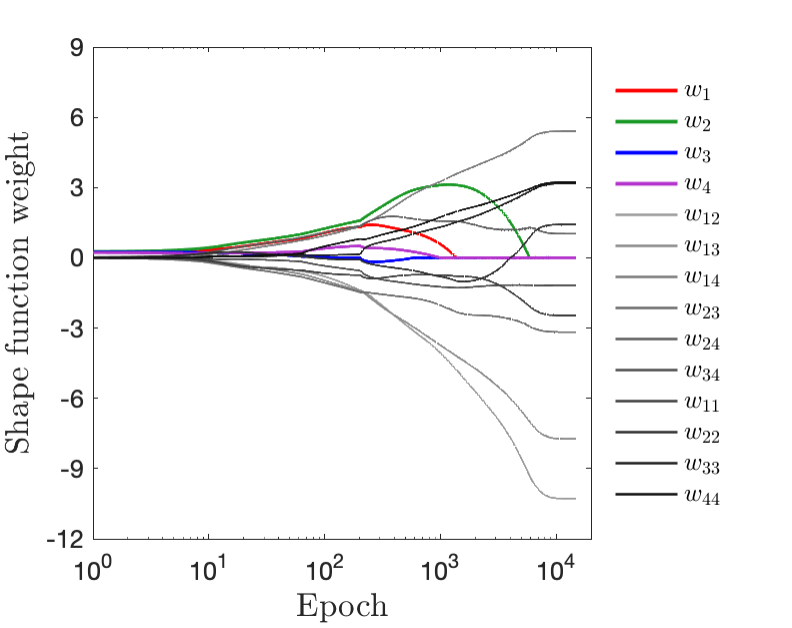}}
\hspace{0.01\textwidth}
\subfigure[]{\label{fig:weight_wo_sparse}
\includegraphics[height=0.375\textwidth]{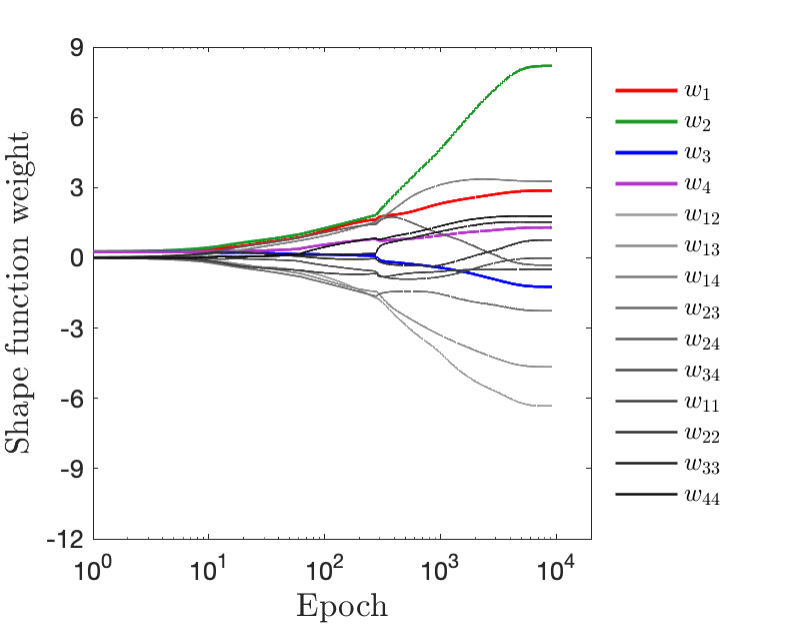}}
\caption{Shape function weights v.s.  epoch: (a) with sparsity control and (b) without. }
\label{fig:sparsity_control}
\end{figure}

Tables \ref{Tab::porous-symb-eqs-sparse} and \ref{Tab::porous-symb-eqs-nosparse} report the corresponding discovered symbolic equations generated by the symbolic regression algorithm. In this study, we deliberately chose the best function with the least loss function to validate our modeling performance for unseen data in the interpolation regime - where data falls inside the convex hull of the training data but was not seen during training. To this end, we plot the yield surface at two different levels in Figure \ref{fig:sparsity_control_yield_surf}. The results suggest that the modeling assumption in Equation \eqref{eq::qnm}, in terms of separability, may be sufficiently robust to avoid overfitting, at least in the interpolation regime.

\begin{table}
  \centering
  \caption{symbolic shape functions found with sparsity promoting loss constraint for porous metal}\vspace{-10pt}
        \begin{tabular}{|p{10cm}|c|c|}
        \hline
        shape function & Complexity score & Loss \\[3mm]
        \hline
        & & \\
        $\begin{aligned} 
        f_1(\bar{\sigma_h}) = \bar{\sigma_h} - \sin{\left(0.07 \left(\bar{\sigma_h} + \exp({\bar{\sigma_h}})\right) \cos{\left(0.96 \bar{\sigma_h} + 0.43 \right)} \right)}
        \end{aligned}$ & 19 & 4.691e-5 \\[5mm]
        $\begin{aligned}
        &f_2(\bar{\sigma}_{vm}) = \bar{\sigma}_{vm} - \left(\bar{\sigma}_{vm} + \cos{\left(\sin{\left(\bar{\sigma}_{vm} \right)} - 0.07 \right)}\right)\\ &\quad \sin{\left(0.15 \sin{\left(\bar{\sigma}_{vm} - 1.01 \right)} \right)}
        \end{aligned}$ & 19 & 4.815e-5 \\[5mm]
        $\begin{aligned}
        &f_3(\bar{\sigma}_{L}) = \bar{\sigma}_{L} \left(\sin(\left(\sin(\left(\sin(\left(0.56 \bar{\sigma}_{L} + 0.56 \cos(\left(\sin( \right.\right.\right.\right.\right.\\
        & \left. \left.\left. \left.\left.  \left(\bar{\sigma}_{L} + \cos(\left(1.0 \sin{\left(\sin{\left(\sin{\left(\bar{\sigma}_{L} \right)} \right)} + 0.99 \right)} \right)) \right)) \right)) \right)) \right)) \right)) - 1.19\right) - 0.34
        \end{aligned}$ & 33 & 2.113e-04 \\[5mm]
        $f_4(\bar{v}) = \begin{aligned}1.04 \bar{v} + \sin{\left(1.41 \exp({- 0.46 \bar{v}}) \sin{\left(\cos{\left(\bar{v} \right)} \right)} \right)} - 0.61\end{aligned}$ & 33 & 5.206e-04 \\[1mm] \hline
        \end{tabular}
\label{Tab::porous-symb-eqs-sparse} 
\end{table}

\begin{table}
  \centering
  \caption{symbolic shape functions found without sparsity promoting loss constraint for porous metal}\vspace{-10pt}
        \begin{tabular}{|p{10cm}|c|c|}
        \hline
        shape function & Complexity score & Loss \\[3mm]
        \hline
        & & \\
        $f_1(\bar{\sigma_h}) = \begin{aligned}\bar{\sigma_h} - 0.29 \sin{\left(\exp({0.76 \bar{\sigma_h}}) \right)} + 0.17\end{aligned}$ & 17 & 4.364e-05 \\[5mm]
        $
        \begin{aligned}
        f_2( \bar{\sigma}_{vm} ) = 
        &\bar{\sigma}_{vm} + \left(1.47 \bar{\sigma}_{vm} + 0.9\right)
        \left(0.04 \cos{\left(\bar{\sigma}_{vm} \right)}
        + 0.04 \cos{\left(\bar{\sigma}_{vm} + 0.9 \right)}\right)
        \end{aligned}$        & 21 & 8.601e-05  \\[5mm] 
        $\begin{aligned}
        f_3(\bar{\sigma}_{L}) = 
        &\left(\bar{\sigma}_{L} \left(\sin{\left(0.48 \bar{\sigma}_{L} \right)} - 0.03 \cos{\left(\bar{\sigma}_{L} \right)} - 1.2\right) - 0.48\right)\\
        & \cos{\left(\sin{\left(\cos{\left(0.58 \bar{\sigma}_{L} + 0.03 \cos{\left(\bar{\sigma}_{L} \right)} \right)} \right)} \right)}\end{aligned}$ & 36 & 2.003e-05 \\[5mm] 
        $\begin{aligned}
        &f_4(\bar{v}) = 
        \bar{v} + \cos(\left(\bar{v} + \cos(\left(\bar{v} \right.\right.\\
         & \left(- 0.57 \bar{v} \cos(\left(\sin(\left(\sin(\left(\cos(\left(2 \bar{v} + \right.\right.\right.\right.\right. \\
         &\left.\left.  \left.\left. \left.\left. \left.  \cos{\left(\bar{v} \cdot \left(0.63 \bar{v} \cos{\left(\sin{\left(\bar{v} \right)} \right)} + 0.63 \cos{\left(\bar{v} \right)}\right) \right)}- \right.\right.\right.\right.\right.\right.\right.\\
          & \left. \left. \left. \left. \left. \left. \left.  0.57 \right)) \right)) \right)) \right)) - 0.57 \cos{\left(\bar{v} \right)}\right) \right)) - 0.57 \right)) - 0.57
        \end{aligned}$ 
        & 42 & 5.971e-04 \\[1mm] \hline
        \end{tabular}
\label{Tab::porous-symb-eqs-nosparse} 
\end{table}

\begin{figure}[h]
\centering
\subfigure[]{\label{fig:bomarito_sparsity_p15}
\includegraphics[height=0.375\textwidth]{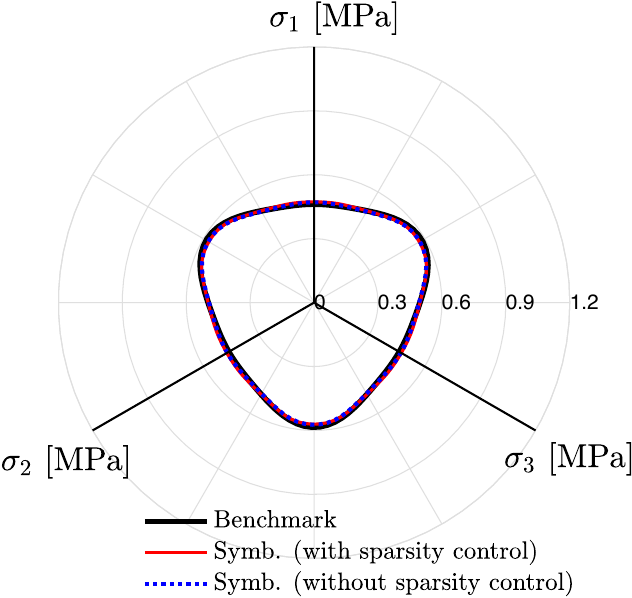}}
\hspace{0.01\textwidth}
\subfigure[]{\label{fig:bomarito_sparsity_p075}
\includegraphics[height=0.375\textwidth]{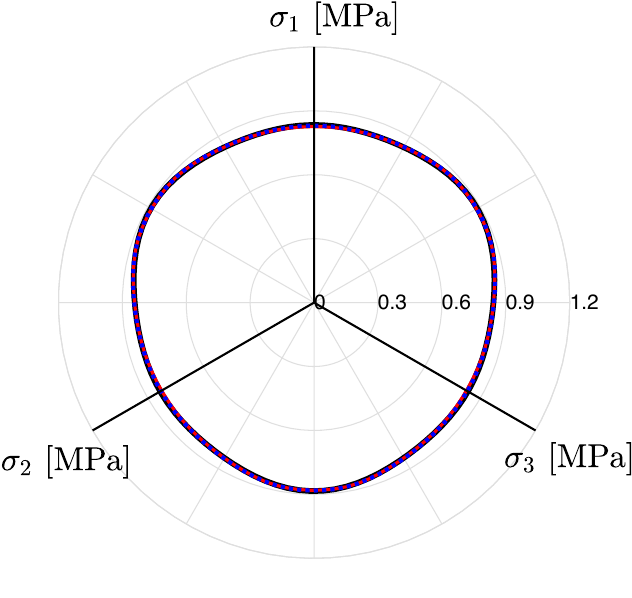}}
\caption{Yield surfaces with and without sparsity control (a) $\bar{\sigma}_h = 1.5$ MPa, $\bar{v} = 0.0645$; (b) $\bar{\sigma}_h = 0.75$ MPa, $\bar{v} = 0.0635$.}
\label{fig:sparsity_control_yield_surf}
\end{figure}

In Section \ref{appendix:symb_reg_comparison}, we conduct a comparison with the brute-force symbolic regression approach directly applied to the data, highlighting the interpretability advantages of our proposed scheme.

\remark[Model and training setup]{
Each utilized MLP consists of one Fourier layer with 20 randomly selected frequencies, followed by three additional hidden layers with 40, 20, and 20 hidden units, respectively. The hidden and output activation layers are \texttt{ReLU} and \texttt{Tanh} layers. Penalty factors are set to $\alpha_{\text{lo}} = 0$ and $\alpha_{\text{ho}} = 0.01$. We set the initial learning rate to 0.005 and continued training for 22,000 epochs. While these hyperparameters were determined through manual trial and error, they are not necessarily optimal.
}

\subsection{Applications in finite element simulations with sybmolic three-invariant plasticity}
\label{sec:FE_simulation}
In this problem, we illustrate how our end-to-end framework can be used to discover symbolic equations for the plastic yield surface, which can then be directly incorporated into finite element simulations. Through this example, we will demonstrate how the QNM approach, with its greater flexibility in modeling assumptions, can lead to more appropriate and simpler symbolic equations compared to the NAM approach.

Our benchmark material model to be replicated via QNM is the Matsuoka-Nakai criterion \citep{matsuoka1974stress}:
\begin{equation}
\label{eq:benchmark_MatsuokaNakai}
f = 
- (I_1 I_2)^{1/3} + (\beta I_3)^{1/3},
\end{equation}
where the stress invariants are defined as: $I_1 = \sigma_1 + \sigma_2 + \sigma_3$, $I_2 = \sigma_1 \sigma_2 + \sigma_2 \sigma_3 + \sigma_3 \sigma_1$, and $I_3 = \sigma_1 \sigma_2 \sigma_3$, while the material parameter $\beta$ depends on the friction angle $\phi_f$:
\begin{equation}
\label{eq:benchmark_MatsuokaNakai_phi}
\beta = \frac{9 - \sin^2\phi_f}{1 - \sin^2\phi_f}.
\end{equation}
By setting the friction angle to be $\phi_f = 30^{\circ}$, we collected a total of 13,200 stress points as a training dataset. 
Specifically, we sampled 20 points along the $p$-axis from 0 to 1,000 MPa, 60 points along the $\theta$-axis from 0 to 2$\pi$, while choosing $N=11$ from 0.85$\rho$ to 1.15$\rho$.  

\begin{figure}[h]
\centering
\subfigure[]{\label{fig:MN_fp}
\includegraphics[height=0.375\textwidth]{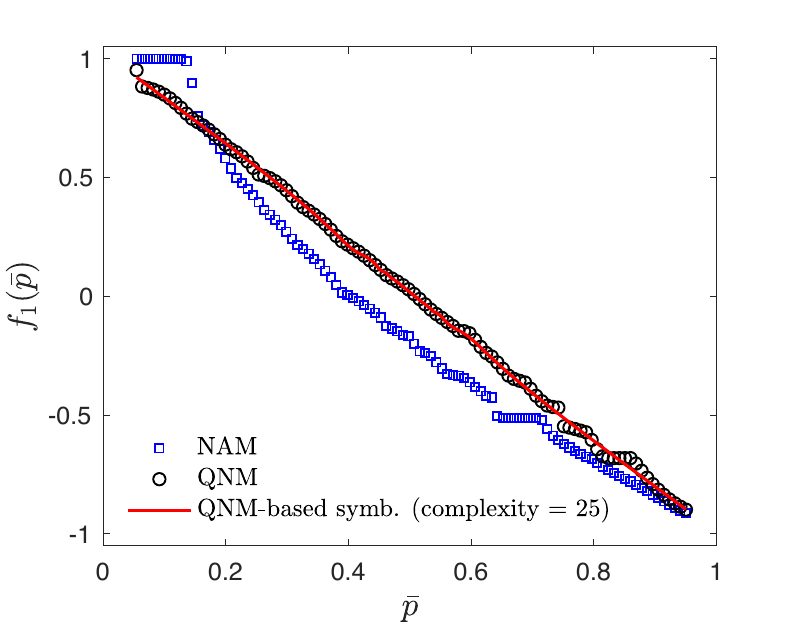}}
\hspace{0.01\textwidth}
\subfigure[]{\label{fig:MN_frho}
\includegraphics[height=0.375\textwidth]{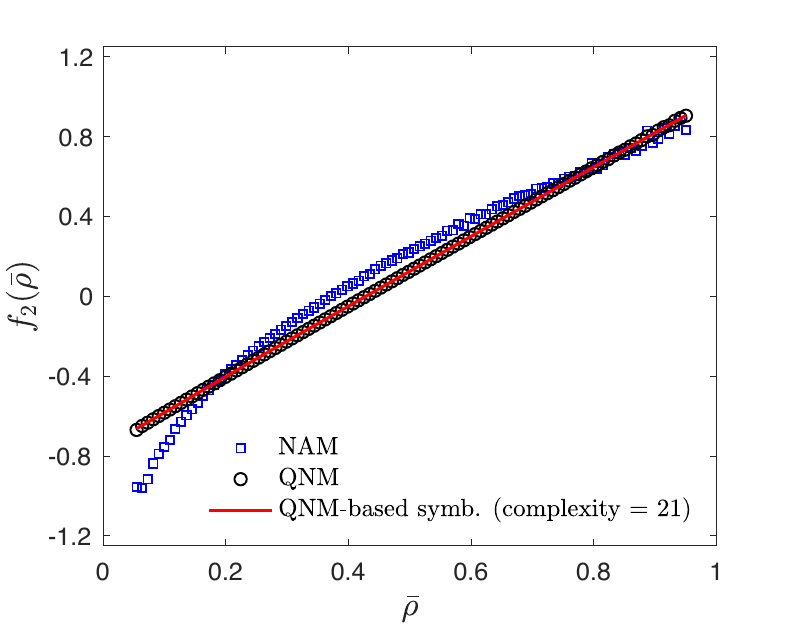}}
\vspace{0.01\textwidth}
\subfigure[]{\label{fig:MN_ftheta}
\includegraphics[height=0.375\textwidth]{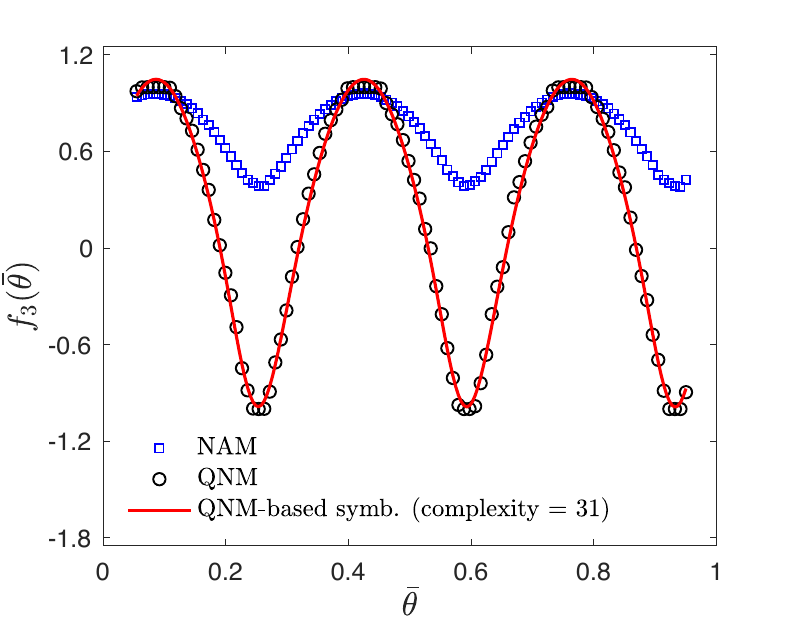}}
\hspace{0.01\textwidth}
\subfigure[]{\label{fig:MN_yield_surf}
\includegraphics[height=0.375\textwidth]{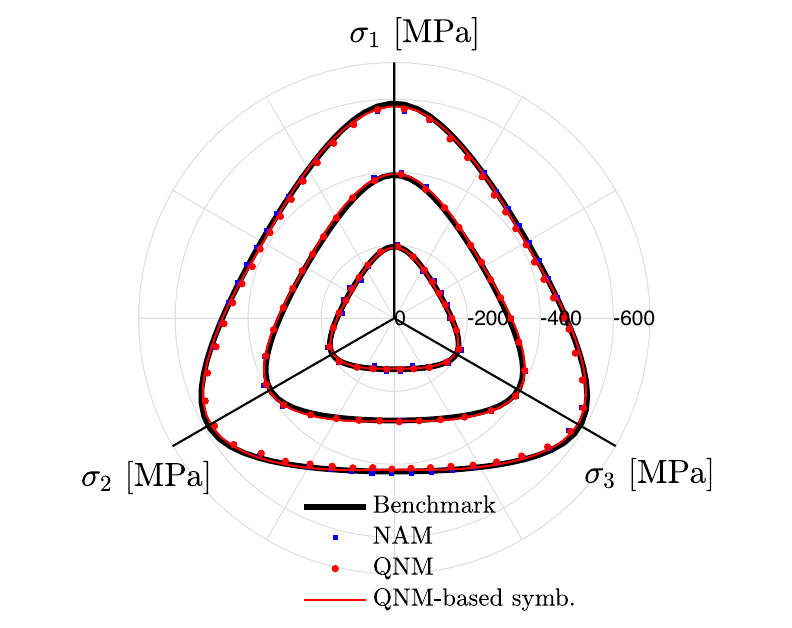}}
\caption{
(a-c) found shape functions for Matsuoka-Nakai data.  (d) found yield surfaces at different confining pressure: $200, 400, 600$ MPa.
}
\label{fig:MN_results}
\end{figure}

The shape functions learned through NAM and QNM are presented in Figs.\ref{fig:MN_fp}-\ref{fig:MN_ftheta}. 
The yield surfaces discovered using these methods, as shown in Fig.\ref{fig:MN_yield_surf}, are in good agreement with the benchmark. However, the shape functions learned through NAM in Figs.~\ref{fig:MN_fp}-\ref{fig:MN_ftheta} exhibit greater complexity and noise, particularly for pressure and radius. This is not surprising, given that the NAM model is unable to account for interactions between input features and, therefore, may increase the complexity of each shape function to improve overall flexibility in capturing the target response.

The learned QNM is expressed as,

\begin{equation}
f = 1.39 f_1(\bar{p}) + 2.18 f_2(\bar{\rho}) + 0.24 f_3(\bar{\theta}) - 0.22 f_1(\bar{p}) f_3(\bar{\theta}).
\label{eq:found-sym-mn}
\end{equation}

All other second-order interactions among the shape functions are nearly zero, except for $f_1(\bar{p}) f_3(\bar{\theta})$. The linear dependence found with pressure (as seen in Fig.~\ref{fig:MN_fp}) and the form of the equation obtained are consistent with the benchmark. This demonstrates the QNM's ability to uncover interpretable relationships among different features and the underlying functional form, which can be useful for the second step of the symbolic regression algorithm.  Table \ref{Tab::theta-MN-symb} summarizes the results of the symbolic regression for the shape function $f_3({\bar{\theta}})$.  The last row in this table is used for the finite element analysis.

\begin{table}
  \centering
  \caption{symbolic shape functions found for $f_3(\bar{\theta})$ in case of pressure-sensitive material}\vspace{-10pt}
        \begin{tabular}{|p{10cm}|c|c|}
        \hline
        Expression & Complexity score & Loss\\[3mm]
         \hline
        & & \\
        $\begin{aligned}
        f_3(\bar{\theta}) = - 1.35 \sin{\left(4.78 \bar{\theta} \right)}\end{aligned}$ & 6 & 5.286e-02\\[5mm]
        $\begin{aligned}
        f_3(\bar{\theta}) = \frac{0.15 - \sin{\left(4.79\bar{\theta} \right)}}{\cos{\left(\cos{\left(2.38 \bar{\theta} - 0.87 \right)} \right)}}\end{aligned}$ & 18 & 1.143e-02 \\[5mm] 
        $\begin{aligned}
        f_3(\bar{\theta}) = \frac{0.13 - \sin{\left(4.79 \bar{\theta} + 6.19 \right)}}{\cos{\left(1.04 \cos{\left(2.38 \bar{\theta} - \cos{\left(\sin{\left(\sin{\left(2.38 \bar{\theta} \right)} \right)} \right)} \right)} \right)}} + 0.03\end{aligned}$ & 31 & 2.928e-03\\[1mm] 
        \hline \end{tabular}
\label{Tab::theta-MN-symb} 
\end{table}

\noindent

We now incorporate the obtained QNM-based symbolic expressions in a boundary value problem solved via the finite element method to showcase the applicability of our proposed approach. 
Specifically, as illustrated in Figure \ref{fig:geometry_and_bcs}, we consider a 20 mm $\times$ 20 mm rectangular plate that is weakened by a circular hole of a radius of 5 mm at its center. 
For simplicity, we limit our attention to a two-dimensional case by assuming plane strain condition while only considering the upper right quarter of our problem domain. 
Our domain of interest is spatially discretized with a mesh that consists of 871 triangular elements that have one integration point each. 
By assuming that our target material behaves linearly in the elastic regime and setting Young's modulus $E = 25$ GPa and Poisson's ratio $\nu = 0.3$, we conduct a finite element simulation under a displacement-controlled regime by prescribing a vertical displacement $\hat{\vec{u}}$ at a rate of $-0.1$ mm/sec on the top, while imposing a 100 MPa compressive traction along the inner radii and the right-hand side of the domain as confinement. 

\begin{figure}[h]
\centering
\includegraphics[height=0.375\textwidth]{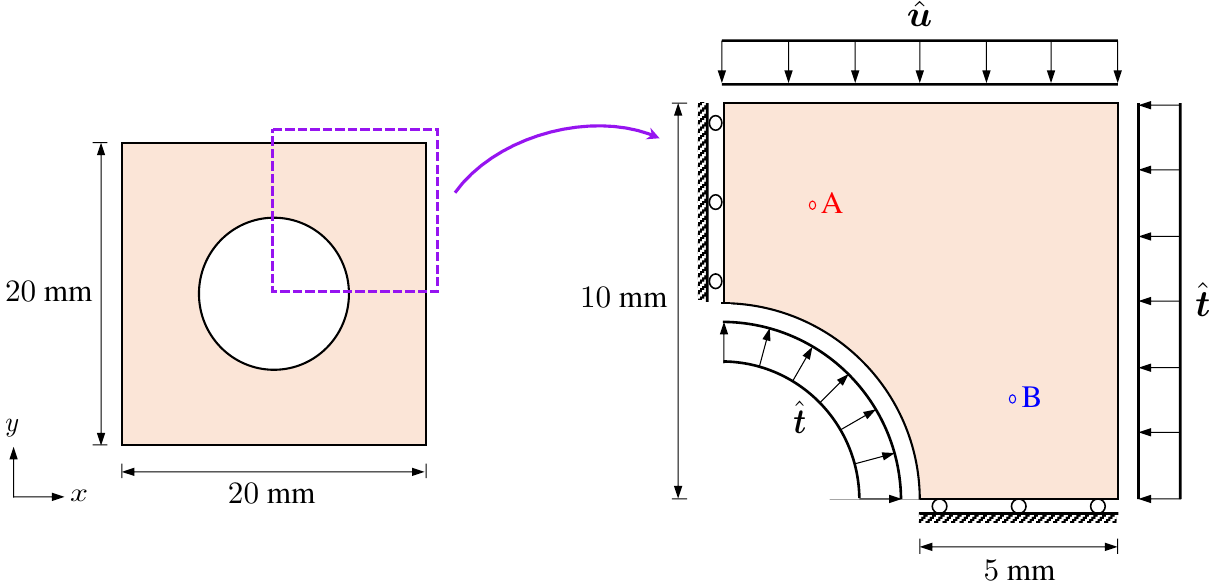}
\caption{Geometry and boundary conditions for the perforated rectangular plate.  The coordinates of points A and B are (2.50, 8.03) and (7.26, 2.47), respectively.
}
\label{fig:geometry_and_bcs}
\end{figure}

Figures \ref{fig:fe_von_mises} and \ref{fig:fe_plas_strain} compare the von Mises stress and the accumulated plastic strain contours obtained from the (a) benchmark and the (b) QNM-based symbolic expressions at $\hat{u}_y = -0.04$ mm, $-0.06$ mm, $-0.08$ mm, and $-0.1$ mm, respectively. 
We observe that the plastic strain first accumulates at the right-hand side of the perforation, where stresses are concentrated and evolves towards the upper right part of the domain of interest, such that it forms a localized pattern. 
Therefore, the stress history recorded at point B near the region where the accumulated plastic strain is localized exhibits a higher level of von Mises stress compared to point A, as illustrated in Figure \ref{fig:stress_evolution}. 
More importantly, the finite element analysis based upon the QNM-based symbolic regression replicates the classical finite element simulation with a benchmark material model, highlighting that our approach is not only capable of discovering the mathematical expression of the yield function from the given set of data without a priori knowledge but also easily replace the constitutive model for continuum-scale simulations.

\begin{figure}[h]
\centering
\includegraphics[height=0.525\textwidth]{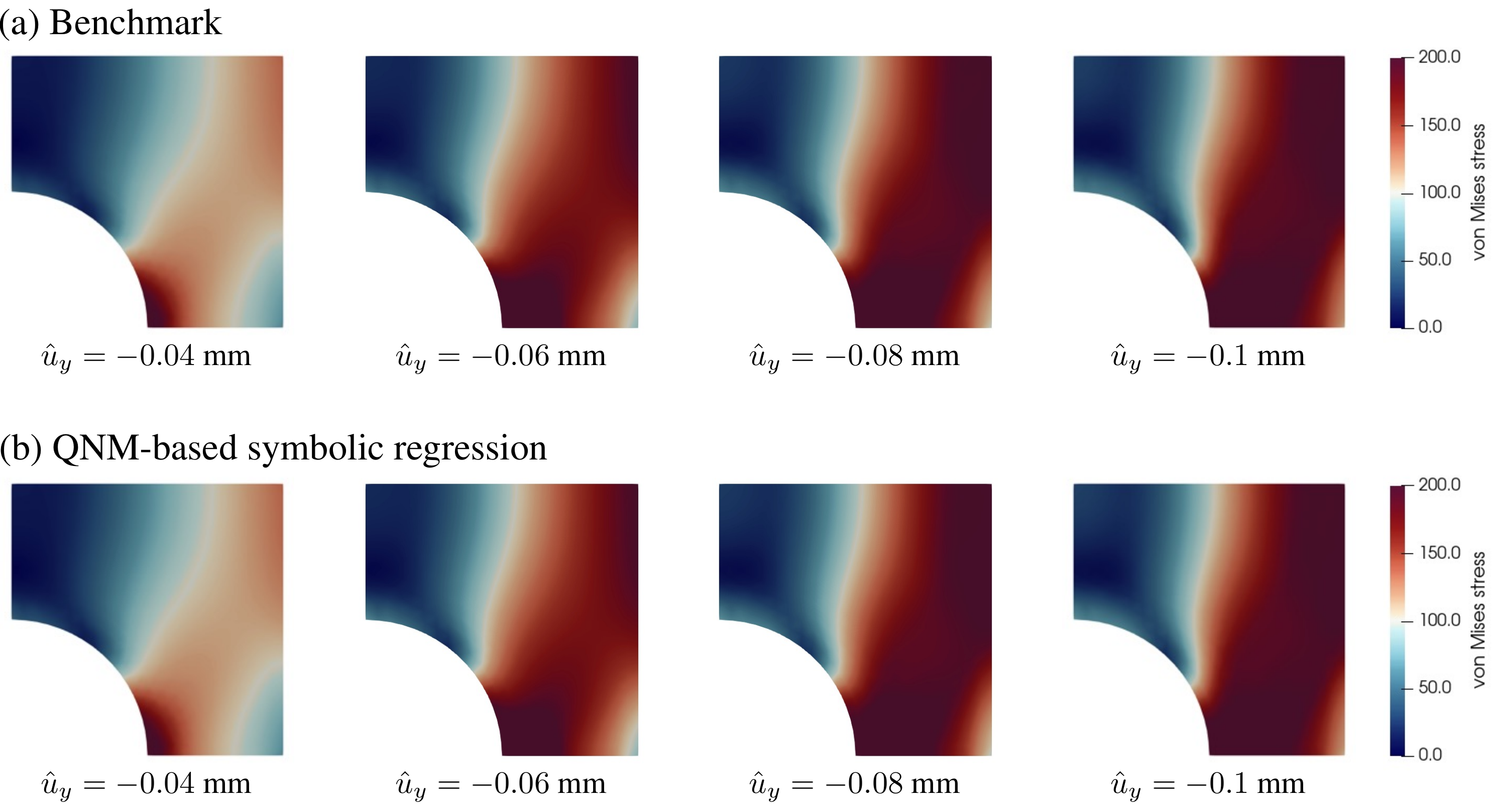}
\caption{
Comparison between the von Mises stress distribution at different stages of loading.
}
\label{fig:fe_von_mises}
\end{figure}

\begin{figure}[h]
\centering
\includegraphics[height=0.525\textwidth]{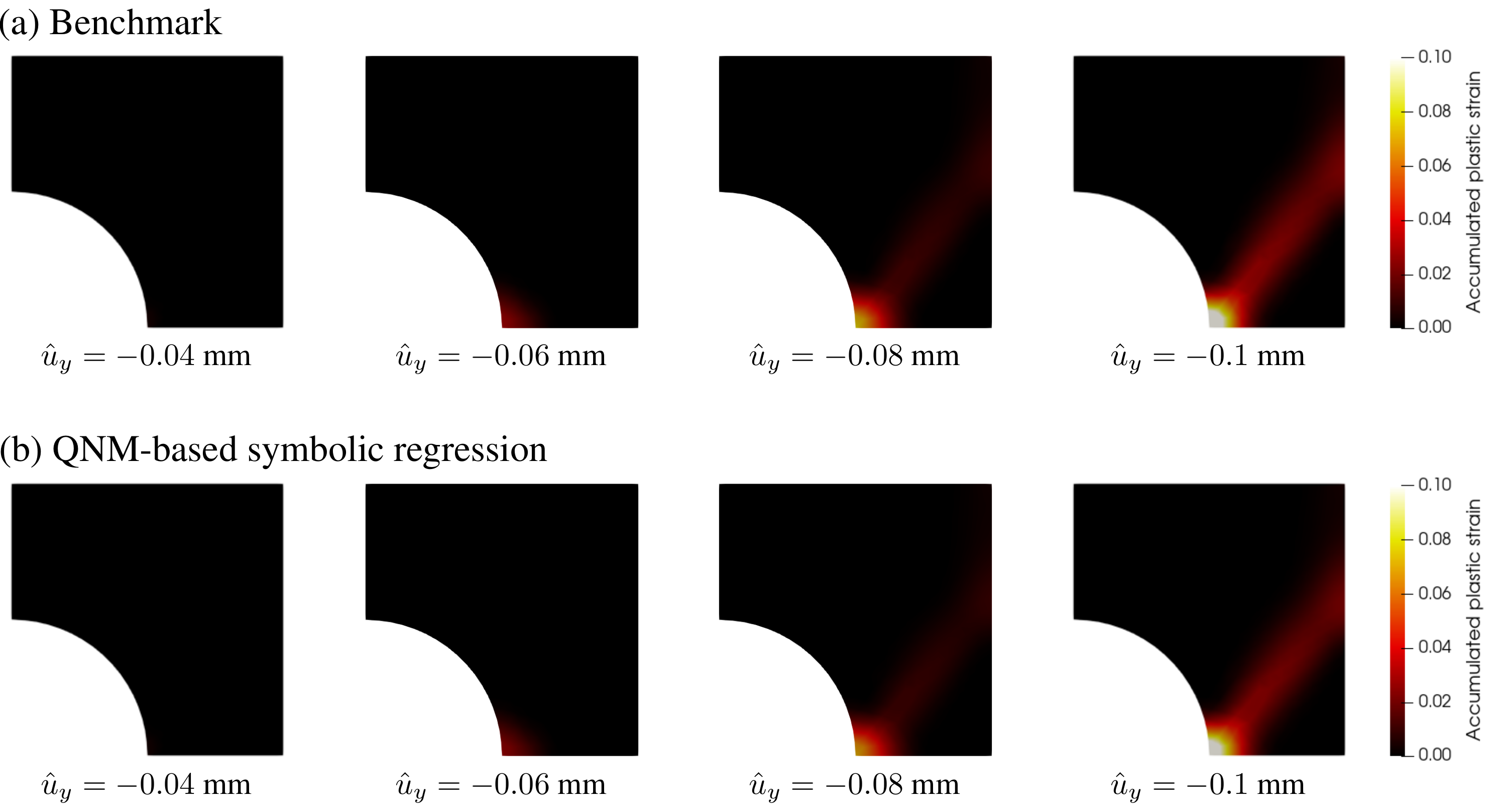}
\caption{
Comparison between the accumulated plastic strain distribution at different stages of loading.
}
\label{fig:fe_plas_strain}
\end{figure}

\begin{figure}[h]
\centering
\includegraphics[height=0.375\textwidth]{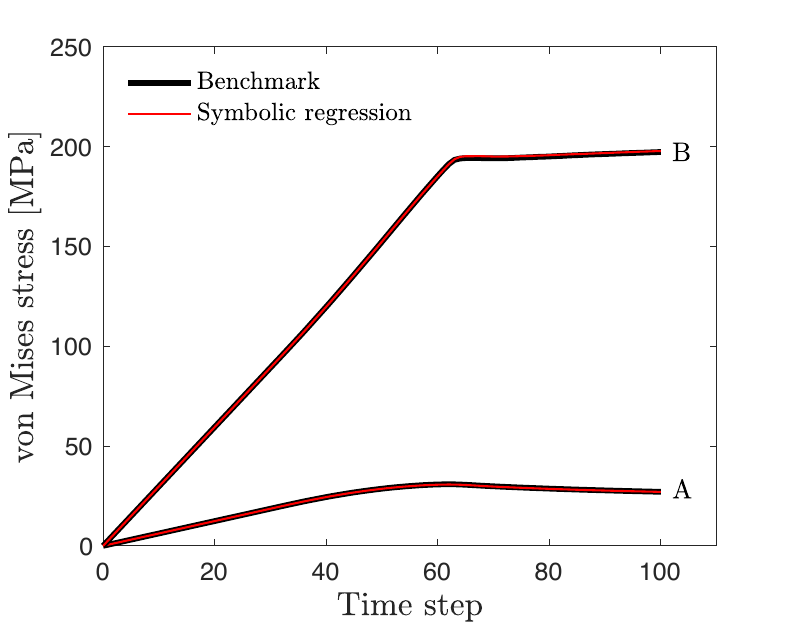}
\caption{Stress evolution in points A and B during the loading.
}
\label{fig:stress_evolution}
\end{figure}

\section{Benchmark of performance}
\label{sec:benchmark}
 In this section, we conduct additional numerical experiments to benchmark the performances of the proposed models against other state-of-the-art approaches in Sections \ref{appendix:symb_reg_comparison} and \ref{appendix:toy_problem}. 
 
\remark{
In Section \ref{appendix:toy_problem},  we compare the performance of NAM and QNM for solving a general regression task with sparse data,  considering the problem dimensionality. ) 
}

 \remark{
Note that the symbolic regression tasks for the discovered shape function are carried out using the \texttt{PySR} package \citep{cranmer2020discovering}. Therefore,  readers should note that the accuracy of learned models obtained from different symbolic regression packages or methods may vary.
}

\remark{For plastic behaviors with discrete mechanisms,  such as the slip system for single crystals,  the yield surface could be non-smooth. 
This non-smoothness cannot be better approximated by increasing the polynomial order in the feature space. 
This limitation of the proposed scheme are discussed in Appendix \ref{appendix:limit}.}

\subsection{Comparisons with the direct symbolic regressions}  
\label{appendix:symb_reg_comparison}
In this study,  we compare the results obtained using our proposed two-step symbolic regression framework to those obtained by applying brute-force single-step symbolic regression directly to the multivariate dataset. 

The total CPU time required to train the QNM is approximately 83 minutes.  Each univariate symbolic regression process took around one minute. Therefore,  the total computational time for our two-step framework is approximately 87 minutes.  In contrast,  when applying the same configuration used for the univariate symbolic regressions to the direct multivariate SR, the SR algorithm takes approximately 43 minutes to find a the following expression,
\begin{equation}
\phi_1 = 1.88 \sin{\tilde{\phi}_1}, 
\end{equation}
where $\tilde{\phi}_1$ reads,
\begin{align}
\begin{split}
\tilde{\phi}_1 = 
    & \bar{\sigma}_{vm} \sin{\left(\sin{\left(\sin{\left(0.41 \bar{\sigma}_{vm} + 0.41 \cos{\left(\sin{\left(\frac{1.08 \sin{\left(0.24 \bar{\sigma}_{vm} \bar{v} \right)} \cos{\left(0.69 \bar{v} \right)}}{\bar{v}} \right)} - 0.16 \right)} \right)} \right)} \right)} \\
    & + \frac{\bar{\sigma}_h + \bar{\sigma}_{vm} + \frac{\bar{v} + \sin{\left(\bar{L} \right)}}{2.85 \bar{v} + 9.4} + \sin{\left(0.22 \sin{\left(\cos{\left(\bar{L} \cos{\left(\cos{\left(\cos{\left(\bar{\sigma}_h + \sin{\left(\bar{L} + 0.86 \right)} \right)} \right)} \right)} \right)} \right)} - 0.33 \right)}}{\cos{\left(\cos{\left(\sin{\left(\cos{\left(\bar{\sigma}_h + 0.73 \right)} \right)} \right)} \right)}}.
\end{split}
\end{align}
%
When the SR algorithm is allocated more time, it discovers $\phi_2 = 1.94 \sin{\tilde{\phi}_2}$ within approximately 3.4 hours where $\tilde{\phi}_2$ reads,
\begin{multline}
    \tilde{\phi}_2 = 
    1.17 (\bar{\sigma}_h + \bar{\sigma}_{vm}) + 0.19 (\bar{v} + \cos(\bar{\sigma}_{vm}) + \cos(\bar{v})) + 0.17 e^{\bar{\sigma}_h} + \sin{\left(\bar{\sigma}_{vm} - 0.95 \right)} +\\
    0.17 \sin{\left(\bar{\sigma}_{vm} + \left(\bar{L} + 1.11\right) \sin{\left(\bar{\sigma}_h + \sin{\left(\cos{\left(\bar{\sigma}_{vm} - 0.71 \right)} \right)} \right)} \right)} +0.17 \sin{\left(\bar{L} + \sin{\left(\cos{\left(\bar{L} + 0.17 \right)} \right)} \right)} - 0.14.
    \end{multline}
The training MSE values for $\phi_1$ and $\phi_2$ are 0.100 and 0.093, respectively.  The RMSE values for random test data (not seen in train data) in each case are 0.0144 and 0.0082, respectively.  However,  RMSE for the proposed two-step method is 0.0070, slightly better than both achieved with less execution time.

$\phi_2$ is more desirable compared to $\phi_1$ in terms of simplicity. However,  both of them may be less desirable in terms of interpretability compared to the proposed two-step framework.  
Since the contribution of each variable in the final yield surface is less apparent.  Additionally,  it is unclear how to simplify this equation and reduce its complexity, which is easily achievable in our framework, as discussed in Section~\ref{sec:low_order_NAM}. 

Figure \ref{fig:direct_comp_yield_surf} demonstrates that QNM-based symbolic regression provides a higher accuracy representation of the target yield surface. This may be due to the proposed divide-and-conquer approach, which has the potential to break down complex learning objectives into simpler ones, possibly resulting in improved learning outcomes.

\begin{figure}[h]
\centering
\subfigure[]{\label{fig:bomarito_direct_vs_qnm_symb_p15}
\includegraphics[height=0.375\textwidth]{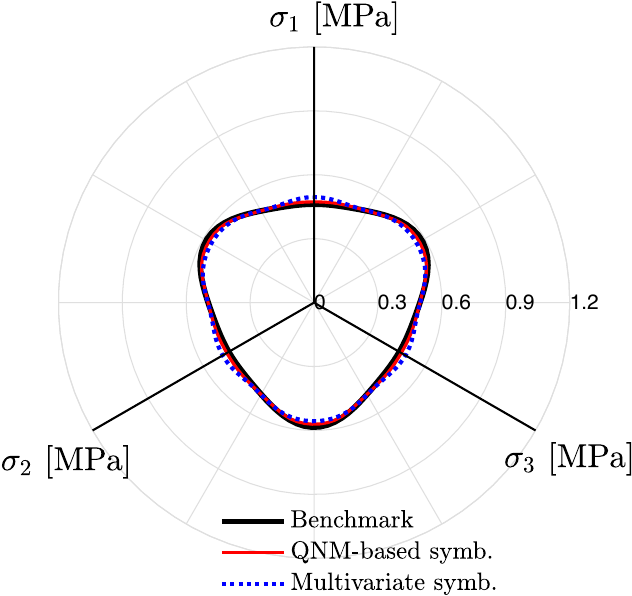}}
\hspace{0.01\textwidth}
\subfigure[]{\label{fig:bomarito_direct_vs_qnm_symb_p075}
\includegraphics[height=0.375\textwidth]{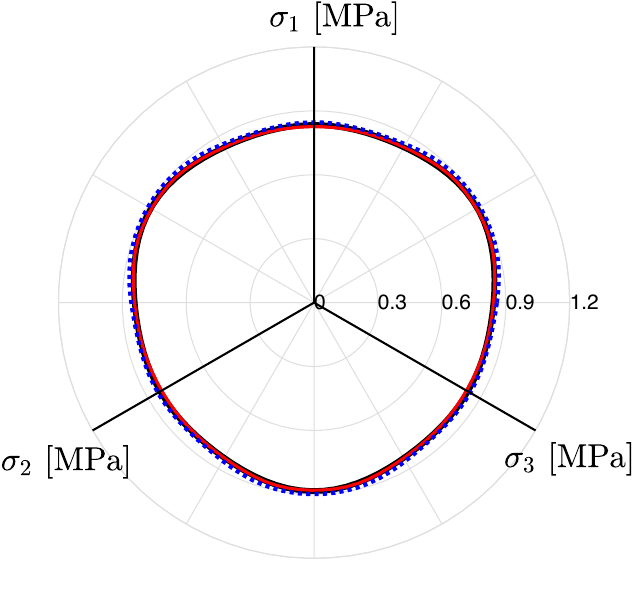}}
\caption{
Yield surface: direct multivariate symbolic vs. QNM-based symbolic regression (a) $\bar{\sigma}_h = 1.5$ MPa, $\bar{v} = 0.0645$; (b) $\bar{\sigma}_h = 0.75$ MPa, $\bar{v} = 0.0635$ -- will work on the captions later. 
}
\label{fig:direct_comp_yield_surf}
\end{figure}

\remark{
Directly comparing the computational time between the proposed method and the direct SR method in this manner may not provide a comprehensive analysis.  It should be noted that the proposed method utilizes both \texttt{PySR} and \texttt{PyTorch}, which are developed by different groups of developers and optimized for different purposes.  On the other hand,  the direct SR method solely relies on \texttt{PySR}. Thus,  due to the differences in the underlying packages and their optimizations,  a direct time comparison may not accurately reflect the performance of each method.
}

\subsection{Comparisons among different methods for sparse data}
\label{appendix:toy_problem}
In this example,  we evaluate the effectiveness of multiple methods for a regression task that involves input features of four dimensions, where the data is relatively sparse.

In this study, we create a regression task with predetermined shape functions, such as polynomial, exponential decay, and multiscale sinusoidal, to assess the method's effectiveness in capturing various shape functions with distinct characteristics.  Additionally, we intentionally exclude one of the input features ($x_4$) in the data generation process to evaluate the method's ability to identify irrelevant features. The data is generated as follows:
\begin{align}
&f_1(x_1) = 3(x_1^3 - x_1), \\
&f_2(x_2)= \frac{1}{x_2 + 1.2}, \\
&f_3(x_3)= 1.5
\left(
-x_3^2 + 0.3 \sin(10\pi x_3) + 0.4
\right
),\\
&f(x_1, x_2, x_3, x_4) = f_1(x_1) + 0.25 f_2(x_2) f_3(x_3) + \mathcal{N}(0, 0.1),
\end{align}
where input variables $x_i$ are sampled randomly from a uniform distribution over the interval $[-1,1]$.  The size of each of the randomly generated training and test datasets is 500 data points.  

Note that the shape function $f_3$ exhibits parabolic behavior at the coarse scale and sinusoidal behavior at the fine scale,  as shown in Fig. \ref{fig:sparse-toy-qnm-shapes}(c).

\begin{figure}[h]
\centering
\subfigure[neural network training]
{\includegraphics[height=0.25\textwidth]{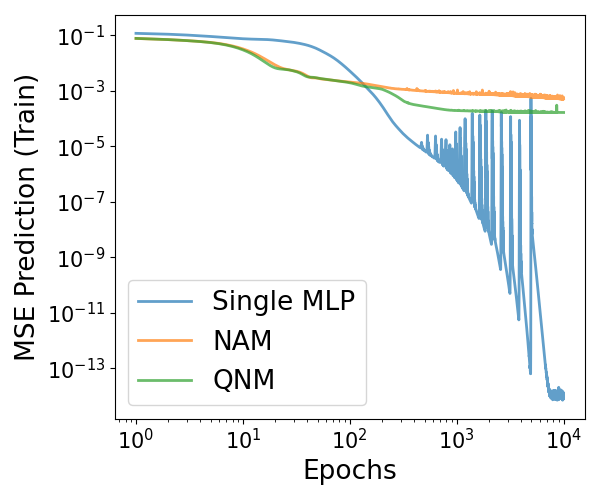}}
\subfigure[residual train data]
{\includegraphics[height=0.25\textwidth]{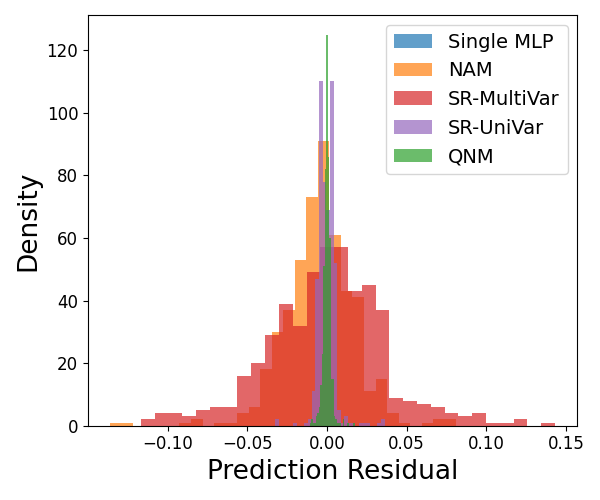}}
\subfigure[residual test data]
{\includegraphics[height=0.25\textwidth]{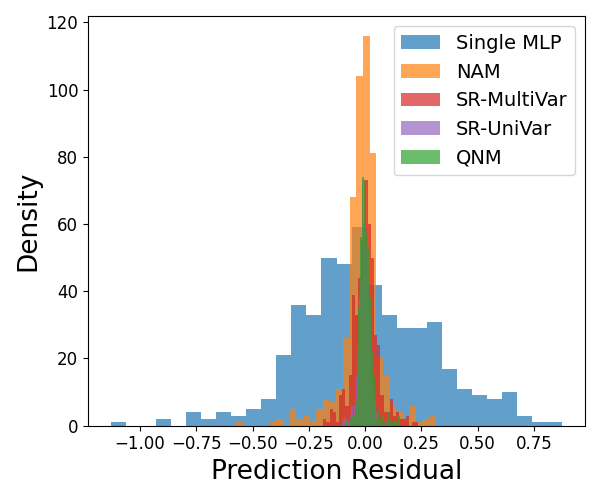}}
\caption{
(a) regression loss values of different models at each training epoch. (b) distribution of the residual error among various models for training data.  (c) distribution of the residual error among various models for test data.  In legends,  SR stands for Symbolic Regression.
}
\label{fig:sparse-toy-errors}
\end{figure}

Figures \ref{fig:sparse-toy-errors}(b-c) display the residuals ($y_{\text{true}} - y_{\text{pred}}$) of the models' predictions using different methods. Figure \ref{fig:sparse-toy-errors}(a) reports the mean squared errors of the predictions during the training process for NAM, QNM, and the vanilla single MLP. The training error for the single MLP method is almost zero,  but it performs poorly on the test data,  as shown in Figure \ref{fig:sparse-toy-errors}(c). This is expected because 500 data points are too sparse for a four-dimensional response surface without any inductive bias. In contrast, NAM and QNM show better generalization than the single MLP since their model assumptions have a more appropriate bias-variance tradeoff and stronger compatibility with the underlying data generation process.
Furthermore,  QNM outperforms NAM in terms of residual errors and train mean squared error, which is expected due to its higher flexibility and structural assumptions fully compatible with the data.   

The quadratic expression of QNM contains non-zero terms, with $w_1 \approx 0.38$ and $w_{23} \approx 0.32$. This means that the structural model discovered by QNM can be written as $\bar{f}_{QNM}(x_1, x_2, x_3) \approx 0.38 \bar{f}_1(x_1) + 0.32 \bar{f}_2(x_2) \bar{f}_3(x_3)$, where $\bar{f}_i$ are learned shape functions. QNM was able to identify the underlying data generation process and discard the irrelevant feature $x_4$.  Interestingly, QNM accurately captured even the complex, multiscale sinusoidal shape function $f_3(x_3)$. While marginal errors can be observed in $f_2(x_2)$, it effectively reflects the exponential decay behavior.

In contrast, NAM identified all terms as non-zero, resulting in the following structural equation: 
\begin{equation}
\bar{f}_{NAM}(x_1, x_2, x_3, x_4) = 0.43 \bar{f}_1(x_1) + 0.51 \bar{f}_2(x_2) + 0.58 \bar{f}_3(x_3) + 0.48 \bar{f}_4(x_4).
\end{equation}
Although NAM and QNM perform similarly in terms of train and test errors, NAM's learned structural equation is misleading. Not only does the irrelevant feature $x_4$ contribute to the model, but its effect is even higher than that of feature $x_1$ ($w_4$ is higher than $w_1$). This can lead to misinterpretation and confusion regarding causality. One possible explanation for this behavior is that, since NAM cannot incorporate interactions among features, it attempts to use $x_4$ as an additional degree of flexibility to minimize prediction loss during training. The learned shape functions are shown in Figure \ref{fig:sparse-toy-nam-shapes}, where only the polynomial shape function $f_1(x_1)$ is discovered by the model.

\begin{figure}[h]
\centering
\subfigure[]
{\includegraphics[height=0.3\textwidth]{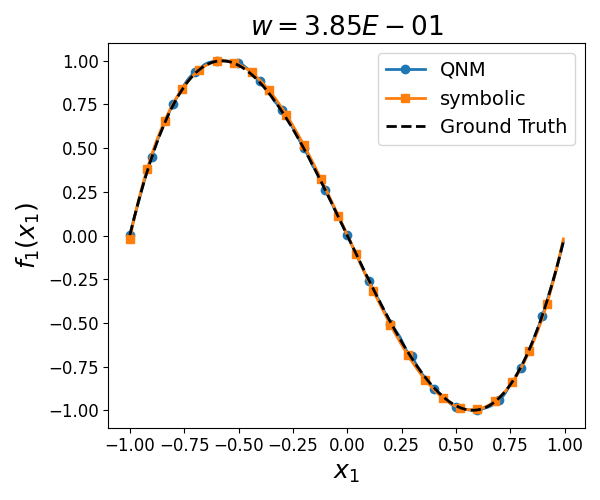}}
\subfigure[]
{\includegraphics[height=0.3\textwidth]{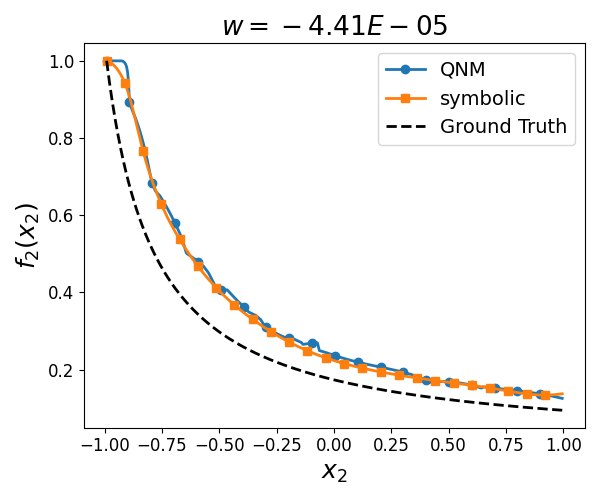}}
\subfigure[]
{\includegraphics[height=0.3\textwidth]{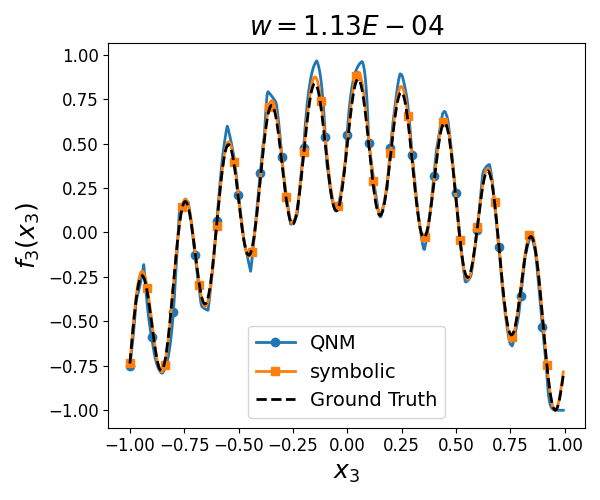}}
\caption{Comparison among the learned shape functions based on the QNM model originally paramtrized via neural networks,  the corresponding symbolic expressions,  and the ground truth. 
}
\label{fig:sparse-toy-qnm-shapes}
\end{figure}

\begin{figure}[h]
\centering
\subfigure[]
{\includegraphics[height=0.3\textwidth]{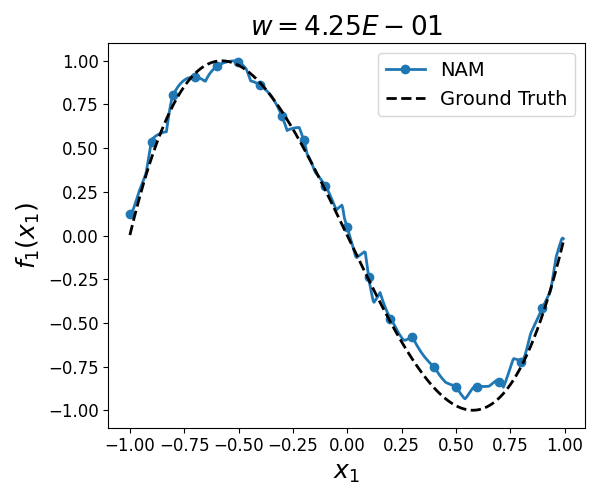}}
\subfigure[]
{\includegraphics[height=0.3\textwidth]{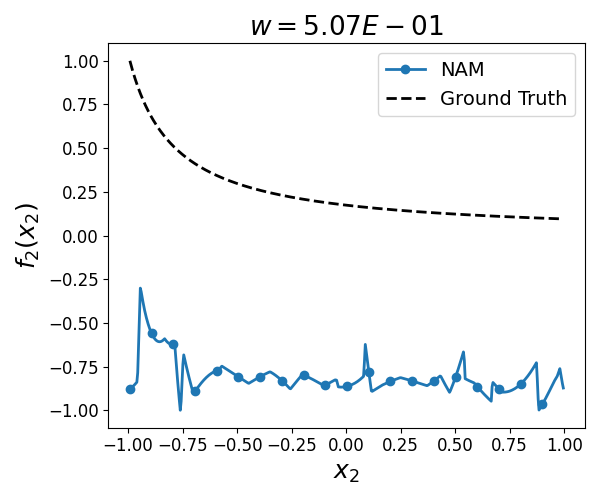}}
\subfigure[]
{\includegraphics[height=0.3\textwidth]{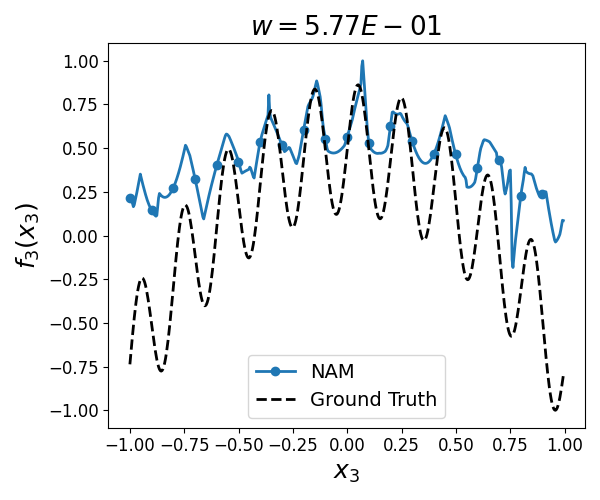}}
\subfigure[]
{\includegraphics[height=0.3\textwidth]{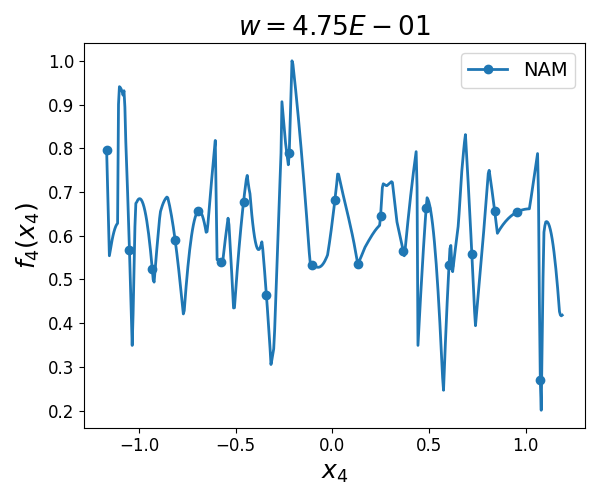}}
\caption{
Comparison between the learned shape functions based on the NAM model and their expected ground truth. 
}
\label{fig:sparse-toy-nam-shapes}
\end{figure}

Figure \ref{fig:sparse-toy-errors} additionally shows the results for two symbolic regression models: ``SR-UniVar'' and ``SR-MultiVar''. The former corresponds to the model obtained by performing symbolic regression on the shape functions learned by the QNM, while the latter is the vanilla multivariate symbolic regression directly performed on the data. While SR-MultiVar does not have the least amount of error in the train data, its performance is comparable to that of SR-UniVar. The symbolic representation discovered by SR-MultiVar is presented below:
\begin{equation}
\frac{- \sin{\left(1.77 x_{1} + 0.17 \right)} + \frac{0.07 e^{- x_{2}} \cos{\left(1.4 x_{3} \right)}}{\sin{\left(\cos{\left(\cos{\left(\frac{e^{x_{3}}}{x_{3}} \right)} - 0.09 \right)} \right)}}}{\cos{\left(\cos{\left(\cos{\left(\sin{\left(x_{1} \right)} \right)} \right)} \right)}}.
\end{equation}
The symbolic representation discovered by SR-MultiVar is fully transparent, and it clearly discards the contribution of the irrelevant feature $x_4$. This is an essential ingredient for model interpretability. However, the equation itself needs to provide an easy way to uncover the underlying data generation process, making it less interpretable than the proposed divide-and-conquer scheme.

\section{Interpretation and validation of yield surface properties}
\label{appx:val-sym-convx}
While it is possible to use numerical tests to test whether the learned plasticity model violates the rules that indicate the loss of desirable properties such as symmetry, convexity, and stability, it is not trivial to prove that the learned model possesses sufficient conditions for these desired properties. Here, we provided selected examples to demonstrate that the interpretability we gained from the multi-dimensional symbolic regression (i.e., the availability of the relatively compact mathematical expression) can be used to establish sufficient conditions mathematically. For brevity, we would not test all the models we have generated in this paper, but the approach we used for post-training analysis is general and should be applicable to other cases.

\subsubsection{Sufficient conditions for convexity}
In this section, we demonstrate how the convexity of the learned models can be rigorously examined analytically.
For brevity, we take the simplest model we obtained in Section \ref{sec:low_order_NAM} in which the yield surface is written as a function of the features, i.e.,  
\begin{equation}
\bar{\phi}(\rho,  \theta) = \rho - c_1\sin(c_2\theta + c_3) - c_4,
\label{eq:yieldexample}
\end{equation}
where $c_1 = 52.73, c_2 = 3.01, c_3 = -9.45, c_4 = 215.01$.  This equation is assembled based on the learned NAM and selecting the first equations in Tables \ref{Tab::petal-symb-eqs-rho} and \ref{Tab::petal-symb-eqs-theta} for radius and angle shape functions, respectively.  Recall that the equations in these tables are provided for the normalized variables.

Note that the availability of the analytical expression of the yield function also enables us to obtain the analytical expression of the plastic flow direction (assuming associative flow rule) and its Hessian.  
Hence, convexity can be analytically established by checking the positivity of the eigenvalues of the Hessian \citep{borja2013plasticity}.
In this case, and thanks to the separability of the discovered equation, we can represent the yield surface (the locus of points that has zero level-set) by writing the radius as a function of angle $\rho(\theta) = c_1\sin(c_2\theta + c_3) + c_4$.  For such representation, the positivity of the Hessian is equivalent to the positivity of the curvature of the polar base vector $\rho(\theta)$\citep{gluge2018does},  
\begin{equation}
\rho^2(\theta)+ 2 \left(\frac{d \rho}{d \theta}\right)^2 - \rho(\theta) \frac{d^2 \rho}{d \theta^2} \ge 0.
\end{equation}
By plugging the symbolic yield surface into the inequality and introducing $X = \sin(c_2\theta + c_3)$, the constraint becomes:
\begin{equation}
 A_1 X^2 + A_2 X + A_3 \ge 0,
\end{equation}
which is a quadratic function of $X$ and $A_1 = c_1^2 (1 - c_2^2)\approx -22410.7$,   $A_2=c_1c_4(2+c_2^2)\approx 125393.6$,
$A_3 = c_4^2 + 2 c_1^2c_2^2 \approx 96611.7$. Since $A_1 < 0$ and the discriminant $\Delta = A_2^2 - 4 A_1 A_3 > 0$, this function is negative unless $X$ is between its two real-valued roots $X_1, X_2 = -0.686,  6.281$.  As such, for 
\begin{equation}
\sin(c_2 \theta + c_3) < -0.686,
\label{eq:convex}
\end{equation}
the yield surface is not convex, 
which is consistent with the visual inspection (see Fig.  \ref{fig:flower_level_set}.)
This inequality,  Eq.  \eqref{eq:convex},  on the other hand, also reveals that the resultant model does not exhibit any spurious high-frequency oscillations, as the curvature of the yield function in Eq. \eqref{eq:yieldexample} evolves smoothly in the parametric space via calculus.

\subsubsection{Symmetry on the $\pi$-plane for yield surface}
One of the commonly shared traits of plasticity models for isotropic materials is the symmetry on the $\pi-$ plane. We have the von Mises plasticity, where the initial yielding is independent of the orientation of the stress path, and the Tresca plasticity in which the yield surface resembles a hexagon. As the $\pi$-plane is obtained by projecting the principal stress space 
onto the plane orthogonal to the hydrostatic axis, the physical implication of symmetry implies the sensitivities of the yielding for different types of shear stress triggered by the following six types of principle stress difference, i.e., $\sigma_{1} - \sigma_{2}$, 
$\sigma_{2} - \sigma_{1}$, $\sigma_{1} - \sigma_{3}$, $\sigma_{3} - \sigma_{1}$, $\sigma_{2} - \sigma_{3}$, and
$\sigma_{3} - \sigma_{2}$.

This symmetry of plasticity can be due to the underlying symmetry of the materials, e.g., the lattice structure of crystals \citep{clayton2010nonlinear}, as well as purposely designed to fulfill specific functions \citep{fleck2010micro} and hence important to preserve in the learned constitutive models. 
The availability of mathematical expression of yield function is helpful for both interpreting and examining the preservation of symmetry, as shown in the following demonstrative example. 
For brevity, consider the simplest symbolic model in Section \ref{sec:low_order_NAM}. 
Since the purpose of this example is to test the expressivity and robustness of the learning algorithm, the training data set used to train the symbolic model is generated by evaluating an already known yield function (Eq. \eqref{eq:benchmark_flower_shape}) with an equally distributed point set in the parametric space. 
As such, it 
possesses rotational symmetry with respect to the following Euler angles along the hydrostatic axis in the principal stress space, i.e., 
\begin{equation}
\theta^{\text{rot-sym}}_n = 2n\pi / k_p,
\end{equation}
 for integer $1\le n \le  k_p$ where $k_p = 3$ in the $\pi$-plane, i.e., the yield surface is invariant under rotations $\theta^{\text{rot-sym}}_n$.  
In the neural network representation, the material symmetry can only be checked through sampling but cannot be proven mathematically. 
In our case, however, the material symmetry can be proven analytically by checking whether $\bar{\phi}(\rho,  \theta)  - \bar{\phi}(\rho,  \theta + \theta^{\text{rot-sym}}_n) = 0$. Just for illustration, we pick one of the found simple models, 
\begin{equation}
\bar{\phi}(\rho,  \theta) = \rho - 52.73\sin(3.01\theta - 9.45) - 215.01,
\label{eq:simplemodel}
\end{equation}
 and check its error term for $n=1$ which corresponds to $\theta^{\text{rot-sym}}_1 = 2\pi / 3$ where $k_p=3$.  For brevity, we introduce $\hat{\theta} = 3.01\theta - 9.45$, then the analytical error term is $\text{err}^{\text{sym}} = - 52.73\left(
\sin(\hat{\theta}) - \sin(\hat{\theta} +  2.006\pi)
\right)$.  
By expanding the second sinusoidal function, we have, 
\begin{equation}
\text{err}^{\text{sym}} = 
- 52.73\left(
\sin(\hat{\theta}) - \cos(2.006\pi) \sin(\hat{\theta}) -\sin(2.006\pi) \cos(\hat{\theta})
\right).  
\end{equation}
As such, the symmetry error $\text{err}^{\text{sym}}$ reads, 
\begin{equation}
\text{err}^{\text{sym}} = -52.73 \left(
0.0002 \sin(\hat{\theta}) - 0.019 \cos(\hat{\theta})
\right).
\end{equation}
This error is maximized at $\hat{\theta} \approx 0.01052$ where the error is $1.00181$. Note that this post-training validation exercise can be carried out easily. Furthermore, a more important lesson is that one may find a remedy to fix the symmetry issue. 

\begin{figure}[h]
\centering
\subfigure[Neural Additive Model]
{\includegraphics[height=0.3\textwidth]{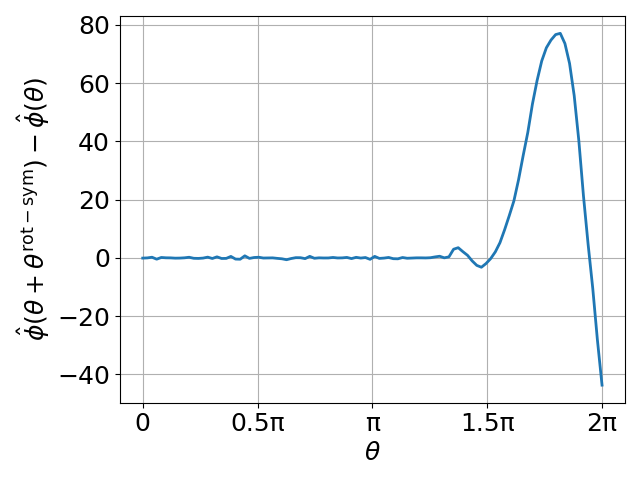}}
\subfigure[Symbolic Model]
{\includegraphics[height=0.3\textwidth]{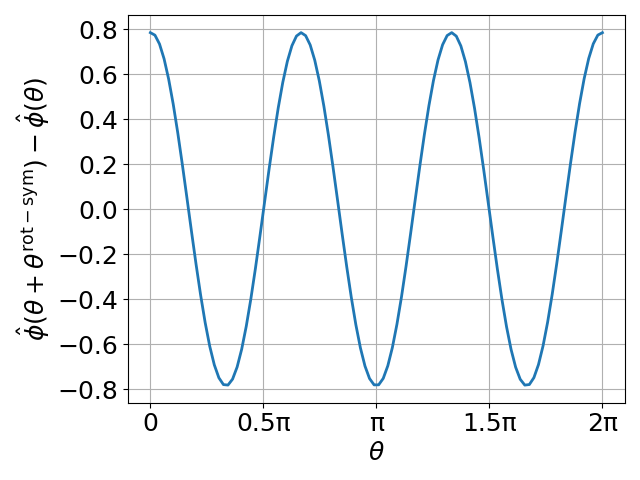}}
\caption{
Symmetry errors of the found yield surfaces in (a) Neural Additive Model (NAM) and (b) Symbolic Model. 
}
\label{fig:rot-sym-check}
\end{figure}

Another interesting effect we found is that the genetic programming used to deduce the analytical expression of the feature shape function may introduce changes in the property of the yield surface even though the difference in the MSE is small. 
Fig. \ref{fig:rot-sym-check} shows the symmetry error of the two models spanned by (a) neural network and (b) symbolic feature shape functions. While both models exhibit errors, the maximum error of the symbolic model is only about $1\%$ of the NAM counterpart. While we should caution against being overoptimistic about the seemingly improved results, the results do suggest that limiting the complexity of the mathematical expression in the symbolic regression process might lead to some regularization effect and filter out undesirable spurious behaviors. However, due to the volatile nature of combinatorial optimization, more research in this area, which is outside of the scope of this paper, is needed before a more definite conclusion can be drawn. 

Symmetry analysis could be more challenging for more complex mathematical expressions, e.g., the model in Section \ref{sec:FE_simulation}. However, leveraging the separation of features, the introduced algorithm significantly reduces complexities that may arise when conducting such analysis for fully connected neural network representations.  Based on the model template found in Eq. \eqref{eq:found-sym-mn}, analyzing symmetry properties in the $\pi$-plane is exclusively associated with the angular shape function $f_3(\bar{\theta})$, which is a univariate function.  Hence,  analyzing the periodicity of functions discovered in Table \ref{Tab::theta-MN-symb} is sufficient for finding the symmetry properties of the entire yield surface.

\section{Conclusions}
\label{sec:conc}
We introduce an integrated framework that combines the expressivity of the neural network and the interpretability of the symbolic regression 
to yield multi-dimensional plasticity models that can be expressed analytically while 
(1) achieving the necessary accuracy for engineering applications
and (2) overcoming the technical barrier of multivariate symbolic regression. 
To strike a balance among the competing objectives of expressivity,  interpretability,  trainability,  and execution speed,  we introduce the following measures. 
\begin{itemize}
\item \textbf{Trainability.} %
To overcome the curse of dimensionality in higher-dimensional symbolic regression problems, we hypothesize that there exists a feature space in which yield surfaces can be expressed as a polynomial function of univariate functions. Each of these functions maps the input variables of the yield surface, i.e., Cauchy stress in our case, to a feature.  The divide-and-conquer nature of the feature generation step allows us to break down a multi-dimensional symbolic regression problem into a set of one-dimensional symbolic regressions, which are easier to solve than their multi-dimensional counterparts.
\item \textbf{Expressivity and accuracy.} We first introduced the QNM architecture,  which generalizes the NAM to incorporate higher-order 
couplings among stress components. This enhanced expressivity is shown to improve the accuracy of the learned model.  Furthermore,  we also adopt  spectral layers (as opposed to the particular activation function used in NAM) to ensure that the resultant neural network architecture is capable of generating univariate shape functions is capable with high-frequency content.
\item \textbf{Interpretability.} The original NAM model relies on the weights of the feature in an additive model to provide interpretability.  We choose a different strategy where interpretability is improved by replacing trained neural network models with symbolic equations. 
A parsimonious loss function is utilized to control the number of higher-order terms in QNM,  promoting simplicity in the final discovered form and preserving the interpretability without significantly comprising the expressivity and accuracy. 
\item \textbf{Execution speed and ease of implementation.} As the constitutive model deduced from machine learning must be applied to a large number of integration points for PDE simulations,  the execution speed of the trained model is crucial for practical purposes.  The availability of analytical expressions with tunable complexity makes the implementation of the machine learning model much easier.  As the symoblic expression does not require the implementation of the neural network,  the resultant model is more portable than the neural network plasticity models.  The relative compact expression (as opposed to the neural network parameterization) also enables us to execute the material subroutine faster and easier to understand.  These features make the resultant models more practical for production. 
\end{itemize}

The proposed machine learning tool is tested against synthetic data 
with a known analytical yield function that is not convex, as well as 
a data set for porous metal with no known analytical solution. 
In all three cases, we found that the proposed method is feasible to train, and the generated model is capable of discovering yield functions with superior accuracy than those obtained from the neural additive model. To ensure third-party validation, the source code is open-sourced.

\section*{Acknowledgments}
\label{sec:acknowledgement}
WCS would like to thank Dr. Sharlotte Kramer and Dr.  Brian Lester for fruitful discussions on implementing neural networks in UMAT that inspire this paper. 
The authors are supported by the National Science Foundation under grant contracts CMMI-1846875 and the Dynamic Materials and Interactions Program from the Air Force Office of Scientific Research under grant contracts FA9550-21-1-0391 with the major equipment supported by FA9550-21-1-0027, with additional funding from the Department of Energy DE-NA0003962 and the UPS Foundation Visiting Professorship at Stanford.   
These supports are gratefully acknowledged. The views and conclusions contained in this document are those of the authors, and should not be interpreted as representing the official policies, either expressed or implied, of the sponsors, including the U.S. Government.  The U.S. Government is authorized to reproduce and distribute reprints for Government purposes notwithstanding any copyright notation herein. The views and conclusions contained in this document are those of the authors, and should not be interpreted as representing the official policies, either expressed or implied, of the sponsors, including the Army Research Laboratory or the U.S. Government. The U.S. Government is authorized to reproduce and distribute reprints for Government purposes notwithstanding any copyright notation herein.

\section*{CRediT authorship contribution statement}
Bahador Bahmani: Conceptualization, Methodology, Software, Validation, 
Formal analysis, Investigation, Data Curation, Writing – Original Draft.  
Hyoung Suk Suh: Conceptualization,  Methodology,  Software,  Validation, 
Formal analysis, Investigation, Data Curation, Writing – Original Draft.  
WaiChing Sun: Conceptualization,  Methodology,  Investigation,  Validation,  Resource,  Writing – Original Draft,  Supervision,  Project administration,  Funding acquisition.

\begin{appendices}
\renewcommand{\theequation}{A\arabic{equation}}
\setcounter{equation}{0}

\section{Limitations of QNM for non-smooth cases}
\label{appendix:limit}
For completeness, we also present a regression example that demonstrates NAM and QNM's inability to achieve the desired level of accuracy.
We synthesized data using the following function:
\begin{equation}
f(x_1, x_2) = |x_1 - x_2| + |x_1 + x_2|,
\end{equation}
This function represents a pyramid, which is depicted in Figure~\ref{fig:failure-data}(a). We obtained prediction results using two models: QNM and NAM. These results are presented in Figures~\ref{fig:failure-data}(b-c). However, since both models rely on restricted modeling assumptions by controlling the amount of possible interactions among input features, they may not be able to represent complex tasks that require higher order interactions among features. To improve the expressivity necessary for this non-smooth learned function, we can extend the QNM to higher-order polynomials or introduce specific enrichment functions (manually or through machine learning) in the feature space to handle the sharp gradient. 
Alternatively, one may also construct feature space locally for coordinate charts that constitute a yielding manifold \citep{xiao2022geometric}. 
These potential improvements are out of the scope of this study but will be considered in the future. 
We included the prediction residuals of these models in Figure \ref{fig:failure-err} for completeness.

\begin{figure}[h]
\centering
\subfigure[train data]
{\includegraphics[height=0.25\textwidth]{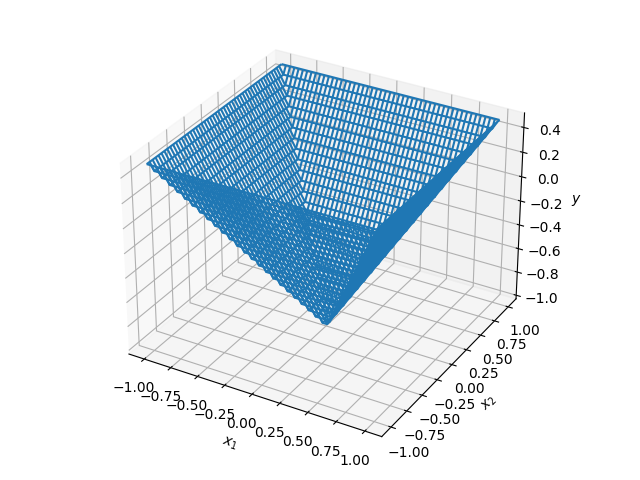}}
\subfigure[QNM prediction]
{\includegraphics[height=0.25\textwidth]{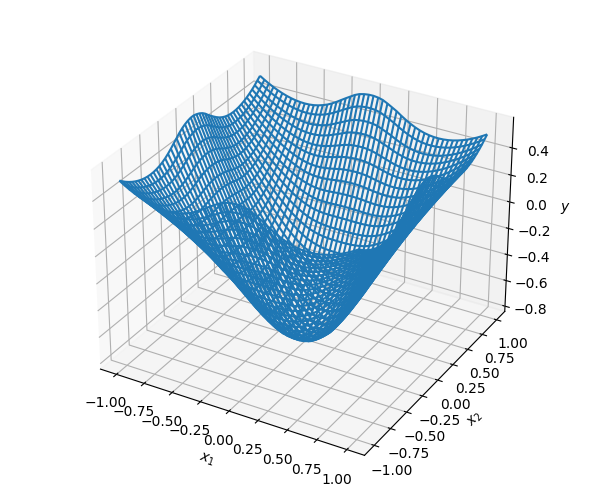}}
\subfigure[NAM prediction]
{\includegraphics[height=0.25\textwidth]{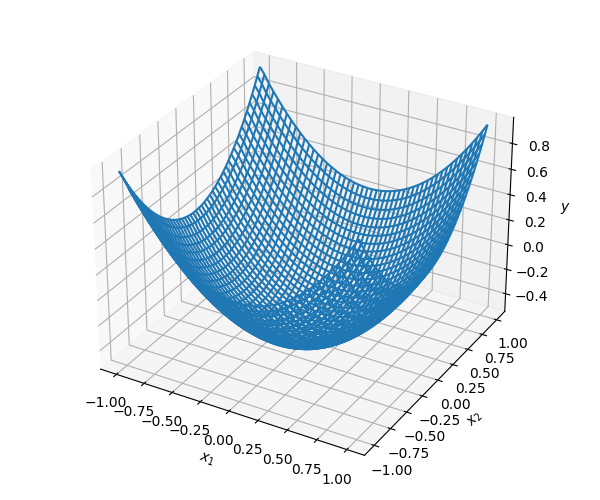}}
\caption{Model predictions in case of (a) the pyramid  function.  QNM and NAM do not achieve satisfactory levels of accuracy.}
\label{fig:failure-data}
\end{figure}

\begin{figure}[h]
\centering
{\includegraphics[height=0.3\textwidth]{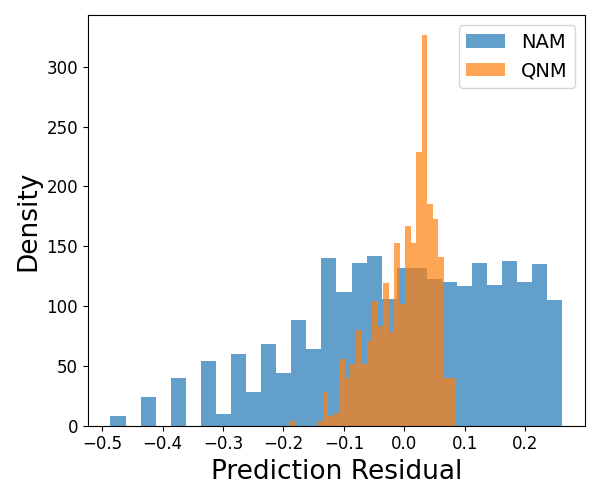}}
\caption{Prediction residuals $y_{true} - y_{pred}$.}
\label{fig:failure-err}
\end{figure}

\end{appendices}

\bibliographystyle{plainnat}

\bibliography{main}

\end{document}